\documentclass{aastex631}
\usepackage{CJK}
\usepackage{xcolor}

\usepackage{longtable}
\usepackage{booktabs, tabularx, multirow, rotating}
\usepackage{natbib}
\hypersetup{hidelinks}
\journalinfo{accepted for publication in ApJ}
\begin{document}
\begin{CJK*}{UTF8}{gkai}

\title{SiO Maser Line-width Diagnostics of AGB Stars and Red Supergiants: The RSG-adjacent Region in the FWZI--Luminosity Plane}

\correspondingauthor{Jun-ichi Nakashima}
\email{junichin@mail.sysu.edu.cn, nakashima.junichi@gmail.com}

\author[0009-0005-7513-9490]{Qiu-Yu Tao (陶秋宇)}
\affiliation{School of Physics and Astronomy, Sun Yat-sen University, Tangjiawan, Zhuhai 519082, Guangdong, P. R. China}

\author[0000-0003-3324-9462]{Jun-ichi Nakashima (中岛淳一)}
\affiliation{School of Physics and Astronomy, Sun Yat-sen University, Tangjiawan, Zhuhai 519082, Guangdong, P. R. China}
\affiliation{CSST Science Center for the Guangdong-Hong Kong-Macau Greater Bay Area, Sun Yat-sen University, 2 Duxue Road, Zhuhai 519082, Guangdong, P. R. China}

\author[0000-0002-1086-7922]{Yong Zhang (张泳)}
\affiliation{School of Physics and Astronomy, Sun Yat-sen University, Tangjiawan, Zhuhai 519082, Guangdong, P. R. China}
\affiliation{CSST Science Center for the Guangdong-Hong Kong-Macau Greater Bay Area, Sun Yat-sen University, 2 Duxue Road, Zhuhai 519082, Guangdong, P. R. China}

\author[0000-0002-3828-9183]{Zehao Zhang (张泽浩)}
\affiliation{School of Physics and Astronomy, Beijing Normal University, Beijing 100875, China}
\affiliation{Institute for Frontiers in Astronomy and Astrophysics, Beijing Normal University, Beijing 102206, China}

\author[0000-0003-1015-2967]{Jia-Yong Xie (谢嘉泳)}
\affiliation{School of Physics and Astronomy, Sun Yat-sen University, Tangjiawan, Zhuhai 519082, Guangdong, P. R. China}

\author[0009-0002-1637-2776]{Chang Liu (刘畅)}
\affiliation{School of Physics and Astronomy, Sun Yat-sen University, Tangjiawan, Zhuhai 519082, Guangdong, P. R. China}

\begin{abstract}
We investigate whether SiO maser line profiles can distinguish asymptotic giant branch (AGB) stars from red supergiants (RSGs) and identify unusual AGB-classified sources near the RSG regime. We compiled 2,956 detected SiO maser spectra for 1,551 objects and constructed reference samples of 1,068 O-rich AGB stars and 36 RSGs. The full width at zero intensity (FWZI) was reliably measured for 1,020 AGB stars and 32 RSGs. AGB stars are concentrated at narrow line widths: 81.5\% have FWZI $<10$~km~s$^{-1}$, whereas 81.2\% of RSGs have FWZI $>15$~km~s$^{-1}$ and none has FWZI $<10$~km~s$^{-1}$. Infrared two-color diagrams provide only partial separation. Combining FWZI with luminosities based primarily on quality-filtered Gaia/Bailer-Jones distances, with independent literature distances for several RSGs, reveals a weak tendency for wider SiO maser lines at higher luminosities but no clean boundary. We therefore define an empirical RSG-adjacent region in the FWZI--luminosity plane, $15 \leq {\rm FWZI} \leq 26.6$~km~s$^{-1}$ and $3.9 \leq \log(L/L_\odot) \leq 4.9$, to characterize the high-luminosity, broad-line tail of the AGB-classified sample. Ten sources fall within this region. Using infrared SEDs, DUSTY fits, OH/H$_2$O maser information, and variability, we find that six are most naturally interpreted as luminous AGB-like objects, two warrant follow-up to assess possible high-initial-mass AGB, RSG-like, or SAGB-related interpretations, and two are less securely constrained because of poor DUSTY fits. The RSG-adjacent region is therefore a physically mixed screening space, not a definitive classification box, for selecting high-priority follow-up targets.
\end{abstract}

\section{Introduction} \label{sec:intro}

Red supergiants (RSGs) and asymptotic giant branch (AGB) stars are the dominant classes of evolved stars known to exhibit circumstellar SiO maser emission \citep{1996A&ARv...7...97H}. Published studies have reported SiO maser emission from roughly 40 Galactic RSGs \citep[see, e.g.,][]{1999PASJ...51..587I,2006ApJ...647L.139N,2010ApJS..188..209K,2012AJ....144..129C,2013AJ....145...22K,Verheyen2012,2012ApJ...760...65F,2010PASJ...62..391D}, as well as from the RSG IRAS 04553$-$6825 in the Large Magellanic Cloud \citep{1996A&A...306L..29V}. SiO maser emission was first detected from the RSG VY CMa in the $J=1-0$, $v=1$ and $v=2$ transitions \citep{1974ApJ...192L..33T,1974ApJ...192L..97B}. Other representative Galactic RSGs with detected SiO maser emission include VX Sgr, NML Cyg, and S Per \citep{2004ApJ...608..480B,2006ApJ...640..982C,2010ApJS..188..209K,2012A&A...544A..42Z}. SiO maser emission has also been detected from RSGs in massive red-supergiant clusters \citep{2006ApJ...647L.139N,2010PASJ...62..391D,Verheyen2012}.

{Existing Galactic SiO maser surveys contain far more detections associated with AGB stars or AGB candidates than with RSGs, but this numerical difference should not be interpreted directly as a difference in the intrinsic SiO-maser occurrence fractions of the two stellar populations.} Among the earliest detections of SiO maser emission from an AGB star was that toward W Hya in the $J=2-1$, $v=1$ transition \citep{1975ApJ...195..359K}. Since the 1990s, large-scale surveys of SiO maser emission at 43 GHz, mainly in the SiO $J=1-0$, $v=1$ and $v=2$ transitions, have been carried out with the Nobeyama 45 m radio telescope toward AGB candidates selected primarily by infrared colors. These surveys covered a wide range of Galactic environments, including the Galactic bulge, the inner Galactic disk, and the outer Galactic disk \citep{1995PASJ...47..815J,1996ApJS..106..463J,1999PASJ...51...95J,1999ApJS..125..257I,2000ApJS..130..351D,2000ApJS..128..571D,2001PASJ...53..305D,2001A&A...376..112I,2003PASJ...55..203N,2003PASJ...55..229N,2004PASJ...56..261D,2004PASJ...56..765D,2006PASJ...58..529F,2007ApJ...664.1130D,2007PASJ...59..559D,2012PASJ...64....4D}. Over roughly two decades, they detected about 2000 circumstellar SiO maser sources.

Large-scale surveys of SiO maser emission at 86 GHz, mainly in the SiO $J=2-1$, $v=1$ transition, have also been conducted with the IRAM 30 m telescope toward infrared-selected late-type and AGB-star candidates, resulting in detections from several hundred sources \citep{2002A&A...393..115M,2004A&A...418..103M,2005A&A...435..575M,2018A&A...619A..35M,2020PASJ...72...63M}. More recently, the Bulge Asymmetries and Dynamical Evolution (BAaDE) survey has searched for SiO maser emission in infrared-selected evolved-star candidates, using the Karl G. Jansky Very Large Array (VLA) for the SiO $J=1-0$, $v=1,2$ transitions and the Atacama Large Millimeter/submillimeter Array (ALMA) for the SiO $J=2-1$, $v=1,2$ transitions. BAaDE has detected SiO maser emission from more than 10,000 infrared-selected evolved-star candidates, most of which are expected to be AGB stars \citep{2015ASPC..497..499S,2019ApJS..244...25S,2020ApJ...892...52L,2024IAUS..380..292S}.

{The numerical dominance of AGB stars among known Galactic SiO maser sources reflects both the underlying population sizes and the observational selection history. The Galactic AGB population is estimated to be of order $10^5$ stars \citep{2002MNRAS.337..749J}. A similarly complete Galactic RSG census is not yet available; for example, a recent Gaia-RVS search identified 6,226 RSG candidates for spectra with S/N $>100$ and 11,539 candidates when the threshold was relaxed to S/N $>50$, illustrating the strong dependence of current RSG counts on selection criteria \citep{2025A&A...694A.152Z}. Reported SiO maser detection fractions are likewise sample dependent. The Nobeyama survey program detected SiO masers from about 2,000 of roughly 3,600 color-selected targets ($\sim56\%$), selected mainly to identify mass-losing O-rich AGB stars and related objects \citep{2007IAUS..242..200D}, whereas an 86~GHz survey of 88 RSGs in massive stellar clusters detected SiO maser emission from 15\% of the sample \citep{Verheyen2012}. Because these surveys differ in target selection, transition, sensitivity, infrared properties, and distance distribution, these values should not be regarded as unbiased Galaxy-wide SiO-maser occurrence fractions. A direct comparison of intrinsic SiO-maser incidence between AGB stars and RSGs remains survey dependent.}

Circumstellar SiO maser lines have been widely used as probes of systemic velocities and gas kinematics in the inner circumstellar envelopes of AGB stars and RSGs. For example, very long baseline interferometry (VLBI) observations by \citet{2003ApJ...599.1372D} mapped the motions of SiO maser spots near the stellar surface to investigate circumstellar-envelope dynamics. In addition, \citet{1990A&A...230..339D,2004PASJ...56..765D,2010PASJ...62..525D,2012PASJ...64....4D} used the line-of-sight velocities of numerous circumstellar SiO maser sources in the Galactic disk and bulge to discuss Galactic dynamics.

Although SiO maser emission from both RSGs and AGB stars has long been observed, their line profiles are empirically known to differ. SiO maser profiles in RSGs tend to be broader than those in AGB stars. Whereas AGB stars commonly show narrow, sharply peaked profiles near the stellar systemic velocity, RSGs often exhibit more complex profiles, including asymmetric extensions and multiple peaks \citep[see, e.g.,][]{2004ApJ...608..480B,2006ApJ...647L.139N,Verheyen2012}.

For example, the SiO maser emission from the archetypal RSG VY CMa has a complex profile with a line width exceeding 40 km s$^{-1}$ \citep[see, e.g.,][]{2013MNRAS.436.1708R}. Similar broad or structured profiles are also seen in other RSGs, such as VX Sgr, NML Cyg, and S Per \citep[see, e.g.,][]{2010ApJS..188..209K}. By contrast, spectra obtained in surveys toward AGB-star candidates, including the Nobeyama SiO maser surveys, generally show narrower profiles, typically with line widths below $\sim$15 km s$^{-1}$, and sharply peaked emission near the stellar systemic velocity \citep[see, e.g.,][]{1999ApJS..125..257I,2000ApJS..130..351D,2000ApJS..128..571D,2003PASJ...55..203N,2003PASJ...55..229N,2004PASJ...56..261D,2004PASJ...56..765D}.

However, circumstellar SiO maser line profiles do not separate cleanly into two distinct categories, namely an AGB-like type and an RSG-like type. The large body of survey spectra accumulated over the past several decades indicates that intermediate profiles are also present. Nevertheless, the relationship between such intermediate or broad-line SiO maser profiles and stellar class has not been investigated systematically in a large and homogeneous framework. One reason is that few studies or databases provide a sufficiently large set of SiO maser spectra with uniform coverage and directly comparable spectral information. In this study, we address this limitation by analyzing a large dataset of circumstellar SiO maser spectra obtained mainly with the Nobeyama 45 m telescope and accumulated over more than a quarter of a century. The dataset contains SiO maser spectra for 1551 objects, including AGB stars and RSGs, and provides a large and relatively homogeneous basis for statistical comparison.

The first aim of this study is to quantify the line-profile properties of RSGs and AGB stars, with particular emphasis on SiO maser line width. We examine whether RSGs have systematically broader SiO maser profiles than AGB stars, as expected from previous observational studies, and assess the overlap between the two populations. We also identify AGB-classified sources that extend toward the broad-line regime occupied by RSGs. In this sense, SiO maser line width is treated here as an empirical diagnostic of evolved-star class, not as a definitive classifier by itself.

The second aim is to combine SiO maser line width with luminosity information in order to examine the high-luminosity and broad-line-width tail of the AGB-classified sample. In the full width at zero intensity (FWZI)--luminosity plane, we define an empirical RSG-adjacent region where AGB-classified SiO maser sources approach the luminosity and line-width regime occupied by RSGs. This region is not intended to define a new stellar class. Rather, it is a practical screening parameter space for selecting unusual AGB-classified SiO maser sources whose physical nature deserves closer investigation.

{One motivation for examining this RSG-adjacent region is that it may isolate sources near the high-initial-mass end of the AGB population, including objects in or near the super-AGB (SAGB) regime. Here, SAGB does not denote a later evolutionary stage reached by an otherwise ordinary low-mass AGB star. Rather, the SAGB designation refers to an initial-mass- and composition-dependent evolutionary channel whose precise mass boundaries depend on metallicity and stellar-model assumptions. SAGB stars ignite carbon off center before entering a thermally pulsing SAGB phase and occupy the transition between the most massive AGB stars and stars that proceed to more massive-star endpoints \citep{2017PASA...34...56D}. Their possible outcomes include oxygen--neon white dwarfs and, over a restricted mass range, electron-capture supernovae. Observationally, SAGB stars are difficult to identify because their high luminosities, cool extended envelopes, and rapid mass loss can make them resemble both massive AGB stars and RSGs; dust-enshrouded OH/IR-like behavior is expected, but is not unique to SAGB stars \citep{2017PASA...34...56D}.}

{This distinction is important because a thick circumstellar shell or high mass-loss rate does not by itself distinguish high initial mass from advanced evolution along the AGB. Relatively low-initial-mass stars can become strongly dust-enshrouded late in their thermally pulsing AGB evolution, whereas higher-initial-mass AGB stars belong to a different underlying stellar population. At a population level, the latter tend to be younger and more concentrated toward the Galactic plane and, in the Galactic disk, are often associated with relatively higher metallicity; in particular, \citet{1994MNRAS.267..711W} showed that Mira kinematics and height above the Galactic plane depend on pulsation period, which is linked to age and initial mass. The observational problem is therefore to distinguish a late-AGB interpretation from a high-initial-mass AGB/SAGB-related interpretation rather than to place all dusty AGB stars on a single evolutionary sequence. In this study, we use the RSG-adjacent region as an empirical guide to examine a heterogeneous set of astrophysically interesting evolved sources and to identify promising targets for future spectroscopic, maser, and molecular-line follow-up observations.}

\section{Datasets and Sample Definition} \label{sec:method}

\subsection{SiO Maser Spectral Data and Ancillary Catalogs} \label{subsec:data-used}

We used observational and archival datasets to construct the SiO maser source sample analyzed in this study. The core dataset consists of SiO maser spectra compiled primarily from 16 published SiO maser line surveys, summarized in Table~\ref{tab:sio_summary}, together with observations of 109 unpublished SiO maser sources presented in Appendix~\ref{app:unpublished_sio_survey}. Most of the published surveys were carried out with the Nobeyama 45-m millimeter-wave radio telescope; the only exception is the survey of \citet{2001PASJ...53..305D}, which was conducted with the Mopra 22-m telescope, as noted in Table~\ref{tab:sio_summary}.

All 16 published surveys and the unpublished Nobeyama observations used single-dish telescopes. The additional RSG SiO maser spectra compiled from the literature were also obtained predominantly with single-dish telescopes, except for two RSGs observed with the Australia Telescope Compact Array (ATCA; \citealt{2012ApJ...760...65F}). Interferometric observations can resolve out spatially extended emission, potentially reducing the measured velocity extent \citep[e.g.,][]{2011MNRAS.415.1083A}. However, excluding these two ATCA sources does not materially change the population-level FWZI difference, as shown in Section~\ref{subsec:FWZI_histogram}.

\begin{figure}
    \centering
    \includegraphics[width=0.72\textwidth]{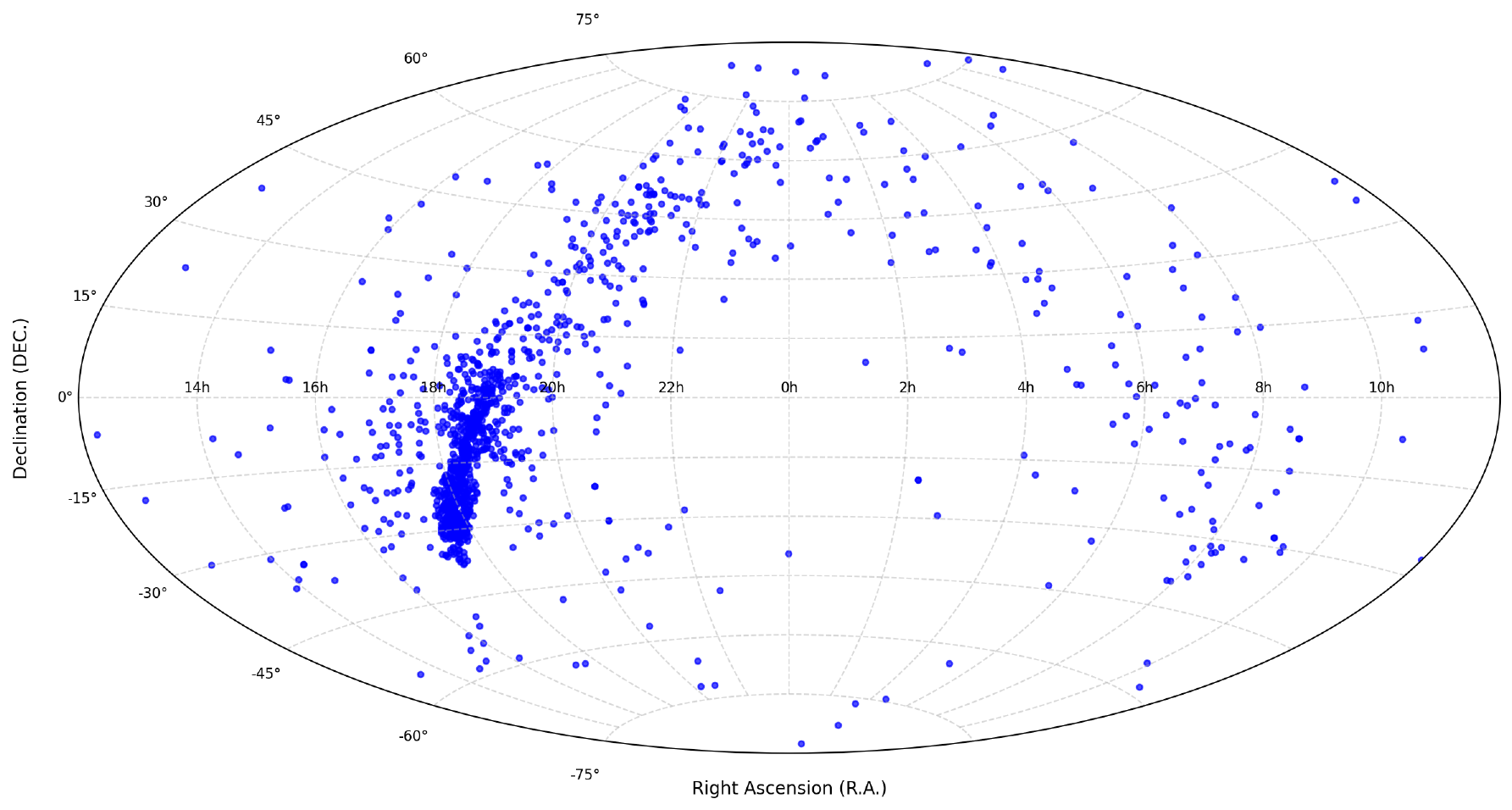}
    \caption{Spatial distribution of the 1,551 SiO maser sources in the compiled dataset, shown in equatorial coordinates using a Mollweide projection. Blue dots indicate individual sources described in Section~\ref{subsec:data-used}.}
    \label{fig:Maser_spatial_distribution}
\end{figure}

Across the published and unpublished datasets, we used 1,472 spectra of the SiO $J=1$--$0$, $v=1$ line (rest frequency: 43.122~GHz), 1,446 spectra of the SiO $J=1$--$0$, $v=2$ line (rest frequency: 42.821~GHz), and 38 spectra of the SiO $J=2$--$1$, $v=1$ line (rest frequency: 86.243~GHz). In total, the dataset contains 2,956 detected SiO maser spectra for 1,551 objects. The spatial distribution of these objects in equatorial coordinates is shown in Figure~\ref{fig:Maser_spatial_distribution}. The survey targets were mainly candidate evolved stars selected from infrared colors, such as the IRAS color $\log_{10}(F_{25}/F_{12})$ and the 2MASS color $H-K_{\rm s}$. In addition to these infrared color-selected targets, known AGB stars \citep{2021ApJS..256...43S} and RSGs in clusters \citep{2006ApJ...647L.139N,2010PASJ...62..391D} were also observed.

Most objects with detected circumstellar SiO maser emission are expected to be AGB stars or RSGs. However, rarer classes of SiO maser sources, such as red-nova remnants, water fountains (WFs), and symbiotic stars, may also be included \citep{2020A&A...638A..17O,2010A&A...510A..69K}. Previous studies based on circumstellar maser catalogs suggest that such atypical objects constitute only a small fraction of the overall circumstellar maser-source population \citep{2007IAUS..242..200D,2024ApJS..270...13F,2025ApJ...978..114X}.

The SiO maser spectra used in this work were provided in ASCII format by the Nobeyama SiO Maser Survey Project team (PI: Shuji Deguchi). The provided dataset includes unpublished SiO maser data for 109 detected sources. Details of this unpublished portion are presented in Appendix~\ref{app:unpublished_sio_survey} with Dr.\ Deguchi's consent. The unpublished observations were carried out with the Nobeyama 45-m telescope from February to May 2011 toward 210 infrared-bright long-period variable stars, of which 109 were detected in SiO maser emission. The observational setup and procedures were essentially the same as those of previous Nobeyama SiO maser surveys \citep[e.g.,][]{2012PASJ...64....4D,2010PASJ...62..525D}. Further details are given in Appendix~\ref{app:unpublished_sio_survey}.

In addition to the SiO maser spectra, we used archival data from the Wide-field Infrared Survey Explorer (WISE) \citep{2010AJ....140.1868W}, the Two Micron All Sky Survey (2MASS) \citep{2006AJ....131.1163S}, the Infrared Astronomical Satellite (IRAS) \citep{1984ApJ...278L...1N}, and Gaia Data Release 3 (DR3) \citep{2023A&A...674A...1G}.

The WISE, 2MASS, and IRAS data were used mainly to examine the infrared properties of the circumstellar SiO maser sources. For WISE, we used photometry in the W1 (3.4~$\mu$m), W2 (4.6~$\mu$m), W3 (12~$\mu$m), and W4 (22~$\mu$m) bands \citep{2010AJ....140.1868W}. To examine mid-infrared variability, we also used multi-epoch W1 and W2 photometric data from the NEOWISE mission \citep{2014ApJ...792...30M}. We retained only WISE measurements with good photometric quality flags (\texttt{ph\_qual = A} or \texttt{B}) and excluded measurements with nonzero contamination flags (\texttt{cc\_flags} $\ne 0$). For 2MASS, we used photometry in the near-infrared $J$ (1.25~$\mu$m), $H$ (1.65~$\mu$m), and $K_{\rm s}$ (2.17~$\mu$m) bands \citep{2006AJ....131.1163S}. For IRAS, we used fluxes in the 12, 25, and 60~$\mu$m bands \citep{1984ApJ...278L...1N}, restricted to measurements with flux-quality indicators of 2 or 3. The corresponding IPAC-hosted dataset DOI citations are the AllWISE Source Catalog \citep{ipac_allwise}, the 2MASS All-Sky Point Source Catalog (PSC) \citep{ipac_2mass_psc}, the IRAS Point Source Catalog v2.1 (PSC) \citep{ipac_iras_psc}, and the NEOWISE-R Single Exposure (L1b) Source Table \citep{ipac_neowiser_l1b}.

For Gaia, we used parallax-based distances and spectra from the Radial Velocity Spectrometer (RVS). The RVS provides medium-dispersion spectra over 845--872~nm, including the Calcium Triplet (CaT) region \citep{2018A&A...616A...6S}. In this work, absorption features in the RVS spectra that are sensitive to effective temperature, surface gravity, and metallicity were used to identify RSG candidates. For parallax-based distances, we adopted the Bayesian distance estimator \texttt{R\_med\_geo} from \citet{2021AJ....161..147B}.

\subsection{Identification of RSGs and AGB Stars in the Dataset} \label{subsec:identify_RSG_AGB}

To compare SiO maser line properties between AGB stars and RSGs, we first constructed reference samples of objects that can be identified with relatively high confidence as either RSGs or AGB stars among the 1,551 objects described in Section~\ref{subsec:data-used}. For comparison, we also compiled from the literature a list of individual RSGs with detected SiO maser emission that are not included among the 1,551 objects.

RSGs with detected SiO maser emission were identified separately for relatively high- and low-mass-loss cases. RSGs with relatively high mass-loss rates ($>10^{-7}\,M_\odot\,\mathrm{yr}^{-1}$) are described in Section~\ref{subsubsec:RSG_high_mass_loss}. Our dataset includes SiO maser observations of RSGs in red supergiant clusters (RSGCs), which are characterized by large infrared excesses and relatively high mass-loss rates. To complement these cluster RSGs, we also compiled individual RSGs with detected SiO maser emission from the literature, including objects that are not part of the 1,551-source dataset described in Section~\ref{subsec:data-used}.

We also investigated whether RSGs with relatively low mass-loss rates ($<10^{-7}\,M_\odot\,\mathrm{yr}^{-1}$) are present in our dataset, as summarized in Section~\ref{subsubsec:RSG_GAIA_RVS}. For AGB stars, we identified counterparts by cross-matching our SiO maser sample with an existing AGB star catalog \citep{2021ApJS..256...43S}, as described in Section~\ref{subsubsec:AGB_identify}.

\subsubsection{RSGs with Relatively High Mass-Loss Rates} \label{subsubsec:RSG_high_mass_loss}

We first describe the cluster RSGs included in the SiO maser spectral dataset. In RSGC1, 14 RSGs have been confirmed \citep{2006ApJ...643.1166F}. Their RSG nature was established from M-type near-infrared spectra and near-infrared absolute magnitudes, based on observations with the 2.1-m and 4-m telescopes at Kitt Peak National Observatory \citep{2006ApJ...643.1166F}. A survey for SiO maser lines ($v=1$ and 2, $J=1$--$0$) toward RSGs in RSGC1 was conducted by \citet{2006ApJ...647L.139N}, resulting in detections from four objects. In Table~\ref{tab:RSG_SiO}, we additionally include one RSGC1 member compiled from a later literature study \citep{Verheyen2012}, resulting in five RSGC1 sources in total.

In RSGC2, 26 RSGs have been identified \citep{2007ApJ...671..781D}. The identification procedure for RSGs in RSGC2 was similar to that used for RSGC1 \citep{2007ApJ...671..781D}. A survey of SiO maser lines ($v=1$ and 2, $J=1$--$0$) toward RSGs in RSGC2 was conducted by \citet{2010PASJ...62..391D}, and SiO maser lines were detected from five stars.

For the RSGCs discovered by Mercer et al., namely [MCM2005b] 4, 8, and 13 \citep{2005ApJ...635..560M}, \citet{2010PASJ...62..391D} also carried out SiO maser observations ($v=1$ and 2, $J=1$--$0$) using the Nobeyama 45-m telescope. They selected RSG candidates using infrared color criteria, namely $F_{12\,\mu{\rm m}} > 1$~Jy in MSX, $K_{\rm s} < 9$, and $H-K_{\rm s} > 0.9$ in 2MASS. In [MCM2005b] 4, six RSG candidates were found, one of which was detected in SiO maser emission. In [MCM2005b] 8, six RSG candidates were found, with SiO maser emission detected from three. In [MCM2005b] 13, six RSG candidates were found, with SiO maser emission detected from two. Thus, across [MCM2005b] 4, 8, and 13, a total of 18 RSG candidates were identified, and SiO maser lines were detected from six of them.

Additional RSGs may also be present among the SiO maser sources outside the known RSGCs. To examine this possibility, we cross-matched the SiO maser sources in our dataset with the RSG catalog of \citet{2024MNRAS.529.3630H}. Through this cross-match, we identified two individual RSG candidates with SiO maser detections: 2MASS J18544742+0025265 \citep{2004PASJ...56..765D} and 2MASS J20211407+3537165 \citep{2005PASJ...57..933D}. The RSG catalog of \citet{2024MNRAS.529.3630H} was constructed using effective temperature and luminosity information, together with comparisons with the RSG locus in the H--R diagram.

For comparison, we also surveyed the literature to compile individual RSGs with detected SiO maser emission that are not represented by spectra in our SiO maser dataset. Well-known RSGs such as VX~Sgr and NML~Cyg were excluded from the target lists of the systematic Nobeyama SiO maser surveys; hence, their spectra are not included in our dataset. We therefore compiled their SiO maser information from the literature. In total, we gathered information for 19 unique RSGs with detected SiO maser emission from six papers: \citet{1999PASJ...51..587I}, \citet{2010ApJS..188..209K}, \citet{2012AJ....144..129C}, \citet{2013AJ....145...22K}, \citet{Verheyen2012}, and \citet{2012ApJ...760...65F}. Among these objects, seven are members of RSG clusters, while the remaining 12 are isolated field RSGs. Additional cluster members compiled from later literature studies are also included in Table~\ref{tab:RSG_SiO}. Thus, Table~\ref{tab:RSG_SiO} contains both RSGs detected in the Nobeyama surveys and RSGs compiled from the literature.

The procedures used to confirm that these objects are RSGs are discussed either in the papers reporting the SiO maser observations themselves or in supporting classification studies, including \citet{2000MNRAS.313..271P}, \citet{2001AJ....121.1050M}, \citet{2005ApJ...628..973L}, \citet{2007A&A...466..151M}, and \citet{2009ApJ...697..701M}.

In total, we compiled 36 RSGs with detected SiO maser emission. Of these, 22 are members of RSGCs and 14 are isolated field stars. The list of these RSGs is provided in Table~\ref{tab:RSG_SiO}.

\subsubsection{Search for Low-Mass-Loss RSGs and Gaia Counterpart Identification} \label{subsubsec:RSG_GAIA_RVS}

We next searched for RSGs with relatively low mass-loss rates ($<10^{-7}\,M_\odot\,\mathrm{yr}^{-1}$) among the 1,551 objects in our dataset, following the method of \citet{2025A&A...694A.152Z}. In general, low-mass-loss RSGs are unlikely to exhibit detectable SiO maser emission. Nevertheless, we carried out this check to identify RSGs in the sample as comprehensively as possible.

Because this search requires Gaia RVS spectra, we also describe here the Gaia counterpart-identification procedure. Gaia-based information, including parallax-based distances and light curves, is used not only for RSG identification but also in subsequent analyses.

\citet{2025A&A...694A.152Z} classified low-mass-loss RSGs into two types by combining the equivalent widths (EWs) of the CaT lines at 8498~\AA{}, 8542~\AA{}, and 8662~\AA{} in Gaia RVS spectra with the infrared color criterion $K_{\rm s}-W3 < 0.5$. Early-type RSGs, with spectral types K0--M2, have relatively high effective temperatures and weak molecular absorption; as a result, the CaT lines dominate and EW(CaT) exceeds 1.1~nm. Late-type RSGs, with spectral types later than M2, show stronger molecular bands and lower effective temperatures, so EW(CaT) decreases below 1.1~nm because of enhanced molecular absorption.

Since the CaT EWs required for this classification are measured from Gaia RVS spectra, we first performed a cross-match between the SiO maser sources and Gaia counterparts. To identify Gaia counterparts, we followed the method adopted by \citet{2024ApJS..270...13F}. Specifically, we first identified WISE counterparts for each SiO maser source and then used the WISE positions as the reference coordinates for the Gaia cross-match. This two-step procedure was adopted because WISE positions are generally more reliable for infrared-bright SiO maser sources than the original coordinates used in the SiO maser observations.

To identify WISE counterparts, we searched the WISE Point Source Catalog within a radius of $20\arcsec$ from each SiO maser position and selected counterparts using both positional information and infrared colors. The adopted infrared color criteria, [W1]$-$[W2] $>-2$ and [W2]$-$[W4] $>-2$, were used to retain sources with colors compatible with cold circumstellar envelopes. Their astrophysical justification has been discussed in detail by \citet{2024ApJS..270...13F} and is not repeated here. Among the WISE sources within the search radius that satisfy these color criteria, we adopted the source closest to the observed SiO maser position as the counterpart.

Next, using the WISE counterpart positions as references, we searched for Gaia counterparts within a radius of $3\arcsec$. For 1,161 sources, a unique Gaia counterpart was found within $3\arcsec$, whereas 364 sources had multiple Gaia candidates within the search radius. When multiple Gaia candidates were present, we adopted the Gaia source closest to the WISE position as the counterpart. If no Gaia counterpart was found within $3\arcsec$, the corresponding SiO maser source was excluded from the subsequent Gaia-based analyses. As a result, 1,525 sources were found to have Gaia counterparts, of which 14 possessed Gaia RVS spectra.

{In subsequent sections, when using Gaia-based distances, we adopted the geometric Bayesian distance estimator \texttt{R\_med\_geo} of \citet{2021AJ....161..147B}. As a quality-control criterion for the Gaia astrometry, we used the Renormalized Unit Weight Error (RUWE), a goodness-of-fit indicator for the Gaia astrometric solution, and required RUWE $<1.4$ \citep{2018A&A...616A...2L}; 909 sources satisfy this criterion. However, RUWE $<1.4$ does not by itself guarantee a reliable Gaia distance for individual AGB stars or RSGs, whose large angular sizes and surface-brightness variations can perturb the astrometric solution \citep{2022A&A...667A..74A}. Accordingly, for five RSGs in the FWZI--luminosity analysis for which independent distances were adopted, we used literature values instead of Gaia/Bailer-Jones distances: VLBI-based distances for VY~CMa, VX~Sgr, S~Per, and NML~Cyg \citep{2019ApJ...885..131R,2018ApJ...859...14X,2010ApJ...721..267A,2012A&A...544A..42Z}, and an association-based literature distance for $\mu$~Cep \citep{2021AJ....162..187M}. In particular, the distance of VX~Sgr used here is the VLBI value rather than its discrepant Gaia/Bailer-Jones estimate. Because Gaia counterpart identification is based on positional matching, possible misidentifications also cannot be completely excluded. These limitations are discussed further in Section~\ref{sec:discussion}.}

We now return to the identification of low-mass-loss RSGs. For RSGs earlier than spectral type M2, molecular absorption is relatively weak and the CaT lines dominate. Consequently, EW(CaT) exceeds 1.1~nm and {is larger than that in red giants or dwarfs}. This behavior can be recognized in EW(CaT)--color diagrams \citep[e.g.,][]{2016A&A...592A..16D}. In the present study, we computed EW(CaT) for 14 sources with Gaia RVS spectra, but none exceeded 1.1~nm. This result suggests that no early-type RSGs are included among these sources.

For RSGs later than spectral type M2, stronger molecular absorption reduces EW(CaT) to values comparable to those of giants, making it difficult to classify a source definitively as an RSG on the basis of EW(CaT) alone. Therefore, \citet{2025A&A...694A.152Z} identified late-type RSGs by progressively removing contaminating objects, such as early-type stars and AGB stars. Because our dataset consists of circumstellar SiO maser sources, the dominant contaminants are expected to be AGB stars. Following their method, we used the color criterion $K_{\rm s}-W3 > 0.5$ to identify AGB-like contaminants. All 14 sources satisfy $K_{\rm s}-W3 > 0.5$, suggesting that they are likely AGB stars.

Consequently, no source in our sample can be securely identified as a low-mass-loss RSG.

\subsubsection{Identification of AGB Stars Based on Existing Catalogs} \label{subsubsec:AGB_identify}

Although most circumstellar SiO maser sources in infrared-selected survey samples are expected to be O-rich AGB stars, RSGs and other evolved or peculiar objects can also be included \citep[e.g.,][]{1996A&ARv...7...97H,2007IAUS..242..200D,2020A&A...638A..17O,2010A&A...510A..69K}. It is therefore necessary to define a reliable AGB subsample for comparison with the RSG sample. For this purpose, we cross-matched our SiO maser source list with the AGB star catalog of \citet{2021ApJS..256...43S}. This catalog contains 11,209 O-rich AGB stars and 7,172 C-rich AGB stars in the Galaxy.

\citet{2021ApJS..256...43S} first identified AGB candidates as Mira-type or semiregular long-period variables using data from the AAVSO and OGLE surveys. These candidates were then cross-matched with infrared catalogs, including the IRAS Point Source Catalog, AKARI \citep{2010A&A...514A...1I}, MSX \citep{2001AJ....121.2819P}, WISE, and 2MASS. The sources were classified into O-rich and C-rich AGB stars based on infrared colors and spectroscopic characteristics, such as the presence or absence of silicate emission at 10 and 18~$\mu$m and SiC emission at 11.3~$\mu$m. \citet{2021ApJS..256...43S} also examined detections and non-detections in previous SiO and OH maser observations to confirm O-rich chemistry in the circumstellar envelopes and to assess evolutionary stages.

We cross-matched our SiO maser sample with the O-rich AGB sample of \citet{2021ApJS..256...43S}. For each SiO maser source, we adopted the WISE counterpart position identified in Section~\ref{subsubsec:RSG_GAIA_RVS} as the reference coordinate, and then performed a $3\arcsec$ cross-match with the catalog of \citet{2021ApJS..256...43S}.

As a result, 1,068 SiO maser sources were matched with O-rich AGB stars in \citet{2021ApJS..256...43S}. In the present study, these 1,068 O-rich AGB stars are adopted as the reliable AGB subsample for the subsequent analyses.

{The available variability and AGB-subclass information is incomplete and heterogeneous across this sample. In addition, the catalog of \citet{2021ApJS..256...43S} and the other catalogs used in this study do not provide a homogeneous, explicit OH/IR-star classification across the sample. We therefore do not subdivide the AGB reference sample into OH/IR, Mira, semiregular, or other subclasses in the subsequent analysis.}

\section{Analysis Methods} \label{sec:analysis_methods}

\subsection{Measurement of SiO Maser Line Widths} \label{subsec:FWZI_Measure}

As described in Section~\ref{sec:intro}, this study examines whether circumstellar SiO maser properties, especially line widths, differ systematically between AGB stars and RSGs. We therefore measured SiO maser line widths for the AGB and RSG samples defined in Section~\ref{subsec:identify_RSG_AGB}. This subsection describes the procedure used to derive those line widths.

We adopted the full width at zero intensity (FWZI) as the line-width measure of each SiO maser profile. The FWZI represents the full velocity range over which emission is detected in the spectrum. It was measured as the velocity difference between the most blueshifted and most redshifted channels containing real emission. The detection threshold was set at $3\sigma$, where $\sigma$ was defined as the root mean square (rms) intensity measured in line-free channels. To distinguish real emission from artificial noise spikes, we required a real signal to consist of at least two consecutive channels with intensities above $3\sigma$; a single channel exceeding $3\sigma$ was regarded as noise.

\begin{figure}
\centering
\includegraphics[width=\textwidth]{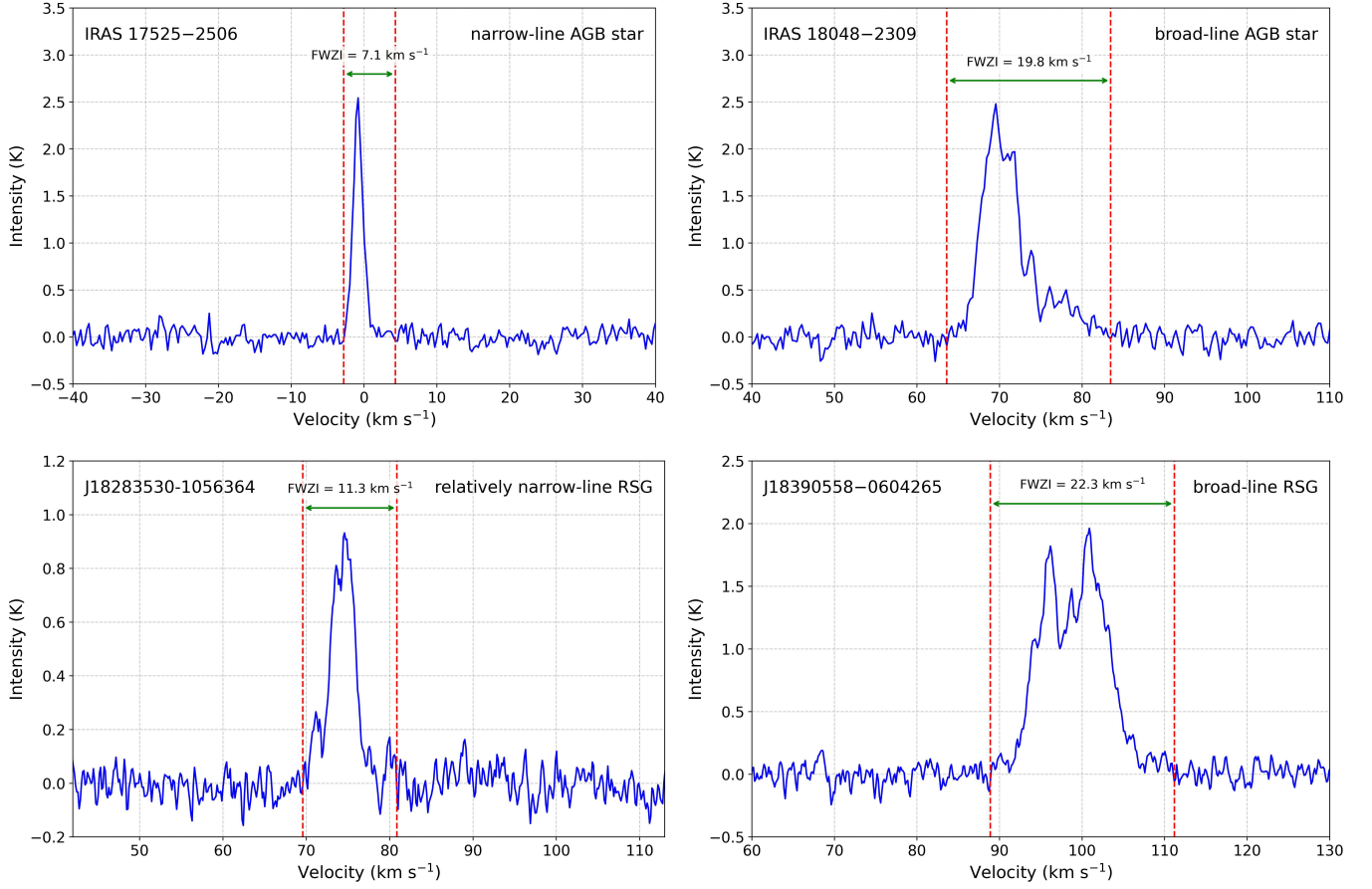}
\caption{Example SiO $J=1-0$, $v=1$ maser spectra illustrating the diversity of line widths observed in AGB stars and RSGs. The four selected sources, from top left to bottom right, are IRAS 17525$-$2506 (narrow-line AGB star), IRAS 18048$-$2309 (broad-line AGB star), 2MASS J18283530$-$1056364 (relatively narrow-line RSG), and 2MASS J18390558$-$0604265 (broad-line RSG), respectively. The horizontal axis represents the Local Standard of Rest (LSR) velocity ($\mathrm{km~s^{-1}}$), and the vertical axis represents the intensity (K). The vertical dashed red lines indicate $V_{\rm blue}$ and $V_{\rm red}$, which were determined using the $3\sigma$ criterion described in Section~\ref{subsec:FWZI_Measure}, requiring at least two consecutive channels to exceed the $3\sigma$ threshold. The FWZI is defined as $\mathrm{FWZI}=V_{\rm red}-V_{\rm blue}$. These spectra were selected not as typical examples of AGB stars and RSGs, but to illustrate the diversity of SiO maser line profiles and the overlap in FWZI between the two populations.}
\label{fig:sio_spec_example}
\end{figure}

Let the velocities at the blueshifted and redshifted edges be $V_{\rm blue}$ and $V_{\rm red}$, respectively. The FWZI is then given by
\begin{equation}
\mathrm{FWZI}=V_{\rm red}-V_{\rm blue}.
\label{eq:fwzi_definition}
\end{equation}
Figure~\ref{fig:sio_spec_example} illustrates this operational definition using representative spectra. The upper two panels show examples of narrow-line and broad-line AGB stars, while the lower two panels show examples of relatively narrow-line and broad-line RSGs. These spectra were selected not as typical examples of AGB stars and RSGs, but to illustrate the diversity of SiO maser line profiles and the overlap in FWZI between the two populations observed in our sample.

{Because the adopted $3\sigma$ criterion is finite, the term FWZI is operational rather than literal in the present analysis: it measures the velocity extent of emission detected above the threshold, not a threshold-independent intrinsic zero-intensity width. Weak line wings may therefore be missed when the rms noise is higher or the maser is fainter, so the measured FWZI can depend on observational sensitivity, source brightness, and indirectly on distance. SiO maser intensity and detailed profile morphology are also known to vary with stellar pulsation phase \citep{2003ApJ...599.1372D,2004A&A...424..145P,2009MNRAS.394...51G,2011MNRAS.415.1083A}. In addition, intrinsic differences among the 43~GHz $v=1$, 43~GHz $v=2$, and 86~GHz $v=1$ masers can introduce transition-dependent selection effects \citep{2018ApJ...862..153S}, while spectral resolution and the number of available transitions can affect the adopted width. We therefore treat FWZI as a threshold-dependent empirical quantity and examine these possible biases in Section~\ref{subsec:FWZI_histogram}.}

\subsection{Luminosity Estimation from Infrared SEDs} \label{subsec:luminosity_estimation}

{To examine the relation between SiO maser line width and luminosity, we estimated luminosities for the subset of sources with adopted distances selected as described in Section~\ref{subsubsec:RSG_GAIA_RVS}. For sources using Gaia astrometry, we required RUWE $<1.4$ and adopted the Bailer-Jones geometric distance \texttt{R\_med\_geo}; for the five RSG exceptions listed in Section~\ref{subsubsec:RSG_GAIA_RVS}, we adopted the corresponding independent literature distances. For each source, the observed spectral energy distribution (SED) was assembled from the infrared photometric data described in Section~\ref{subsec:data-used}. The luminosity was calculated by integrating the observed SED,}
\begin{equation}
L = 4\pi d^{2} \int F_\lambda \, d\lambda,
\label{eq:luminosity}
\end{equation}
{where $d$ denotes the adopted distance defined above, and the integration was performed using the trapezoidal rule. The resulting luminosities are used in Section~\ref{subsec:Final_Candidate} to construct the FWZI--luminosity diagram.}

\section{Results}\label{sec:Result}

\subsection{SiO Maser Line-Width Distributions of AGB Stars and RSGs}
\label{subsec:FWZI_histogram}

\begin{figure}
    \centering
    \includegraphics[width=\textwidth]{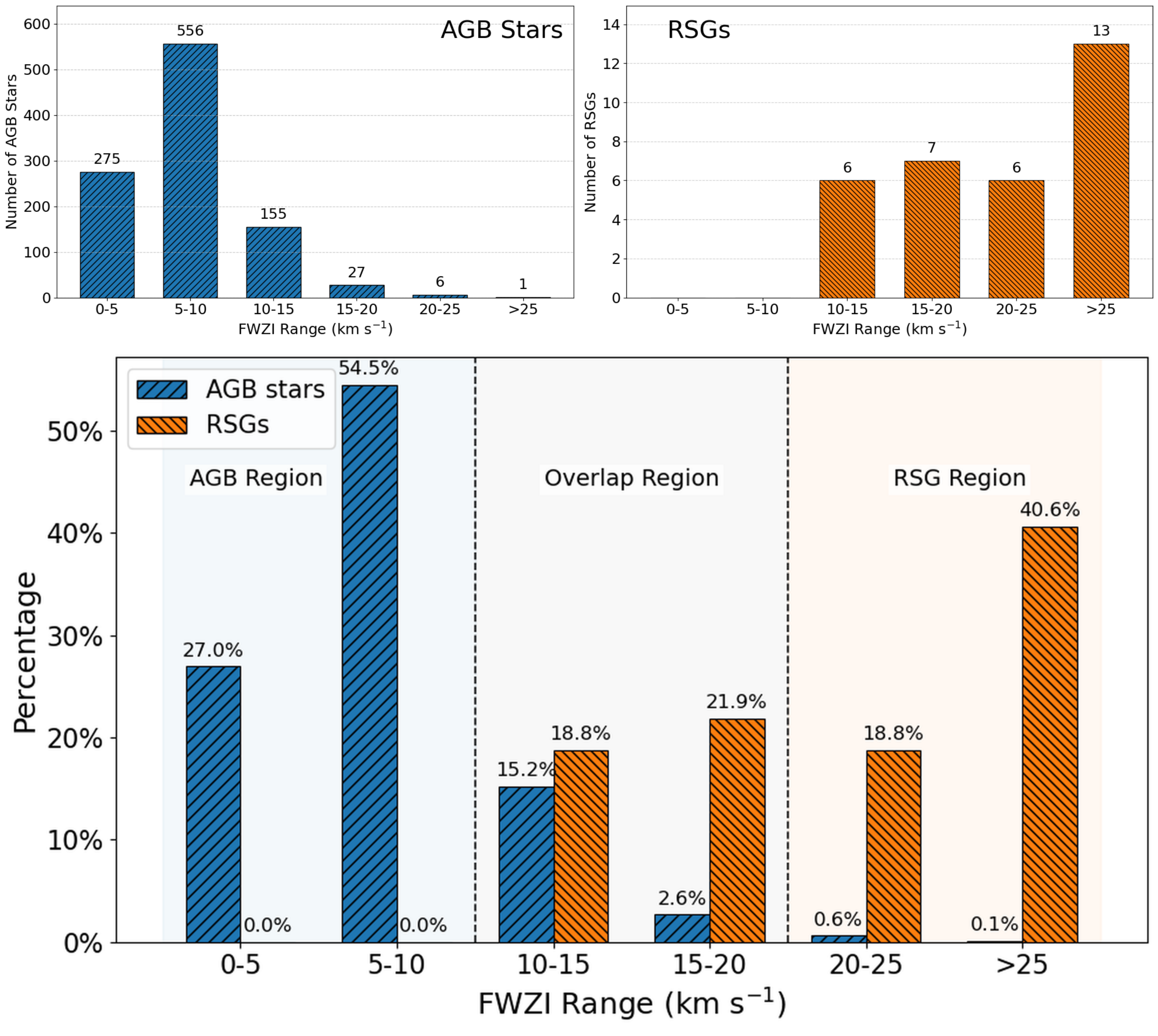}
    \caption{Comparison of the SiO maser line-width (FWZI) distributions of the AGB and RSG subsamples for which FWZI could be reliably measured (1,020 AGB stars and 32 RSGs). The two upper panels show the source counts in each FWZI bin for the AGB (left, blue) and RSG (right, orange) samples, while the larger lower panel shows the corresponding percentage distributions normalized by sample size for direct comparison. The FWZI axis is divided into bins of 0--5, 5--10, 10--15, 15--20, 20--25, and $>25$~km~s$^{-1}$. For convenience, three empirical line-width regions are defined for the present sample: the ``AGB Region'' (FWZI $<10$~km~s$^{-1}$), the ``Overlap Region'' (10--20~km~s$^{-1}$), and the ``RSG Region'' (FWZI $>20$~km~s$^{-1}$). Here, the ``Overlap Region'' denotes the central range in which the two classes overlap most prominently, although the two distributions also overlap in the 20--25 and $>25$~km~s$^{-1}$ bins. Because the RSG percentages are based on only 32 objects, they are subject to relatively large statistical uncertainties.}
    \label{fig:FWZI_histogram}
\end{figure}

We measured the FWZI of each SiO maser spectrum following the procedure described in Section~\ref{subsec:FWZI_Measure}. For sources detected in multiple transitions, we adopted the largest measured FWZI. Figure~\ref{fig:FWZI_histogram} compares the FWZI distributions of the AGB and RSG subsamples for which reliable measurements were obtained, comprising 1020 AGB stars and 32 RSGs. The two upper panels show the source counts for the AGB and RSG samples, and the lower panel shows the corresponding distributions normalized by sample size. The FWZI bins are 0--5, 5--10, 10--15, 15--20, 20--25, and $>$25~km~s$^{-1}$. For descriptive purposes, we define three empirical regimes: the AGB region (FWZI $<10$~km~s$^{-1}$), the overlap region (10--20~km~s$^{-1}$), and the RSG region (FWZI $>20$~km~s$^{-1}$). The strongest overlap occurs at 10--20~km~s$^{-1}$, although the distributions also overlap in the 20--25 and $>$25~km~s$^{-1}$ bins.

The two samples have markedly different FWZI distributions. AGB stars are strongly concentrated at narrow widths: 81.5\% have FWZI $<10$~km~s$^{-1}$, and only 3.3\% have FWZI $>15$~km~s$^{-1}$. Only one AGB star lies in the $>$25~km~s$^{-1}$ bin, with a maximum FWZI of 26.6~km~s$^{-1}$. By contrast, 18.8\% of the RSGs lie in the 10--15~km~s$^{-1}$ bin, and 81.2\% have FWZI $>15$~km~s$^{-1}$. No RSG in the present sample has FWZI $<10$~km~s$^{-1}$. Thus, AGB stars are strongly concentrated at low FWZI values, whereas the RSG distribution is shifted toward broader profiles.

{FWZI is therefore an asymmetric empirical discriminator. A narrow profile strongly favors an AGB classification, whereas a broad profile does not uniquely indicate an RSG because a small number of broad-line AGB stars are present. We performed several checks to assess whether the population-level difference could instead be produced by measurement or selection effects. First, FWZI shows no statistically significant correlation with adopted distance for the full distance sample ($n=335$, $r=-0.043$, $p=0.432$), the AGB subsample ($n=309$, $r=-0.020$, $p=0.729$), or the RSG subsample ($n=26$, $r=-0.359$, $p=0.072$), although the RSG sample is small. Sensitivity and source brightness do have a measurable effect. For the 43~GHz $v=1$ data, the AGB FWZI correlates weakly with rms noise ($r=0.110$, $p=0.0001$) and peak intensity ($r=0.224$, $p<0.0001$); for RSGs, the FWZI--rms correlation is not significant ($r=0.142$, $p=0.464$), while the correlation with peak intensity is stronger ($r=0.573$, $p=0.001$). These trends are consistent with faint wings being more readily detected in brighter spectra. Nevertheless, in a high-S/N subset with S/N $>20$, 80.6\% of the 103 AGB stars have FWZI $\leq15$~km~s$^{-1}$, whereas all 22 RSGs have FWZI $>15$~km~s$^{-1}$. Thus, sensitivity and brightness can modify individual FWZI measurements and the detailed distribution, but the empirical AGB--RSG difference remains present.}

{We also tested transition-dependent and multiple-measurement effects. The availability of the 43~GHz $v=1$ transition is similar for AGB stars and RSGs (85.2\% and 87.5\%, respectively), whereas the 43~GHz $v=2$ and 86~GHz $v=1$ fractions are not identical between the two classes. However, the qualitative difference between the narrow AGB and broad RSG distributions persists when the samples are restricted separately to 43~GHz data, to 43~GHz $v=1$, or to 43~GHz $v=2$. Sources with multiple valid transitions have larger median adopted FWZI than sources with only one valid transition (7.8 versus 5.0~km~s$^{-1}$ for AGB stars and 23.6 versus 17.5~km~s$^{-1}$ for RSGs), showing that the availability of multiple lines is itself associated with source properties and/or detectability. Replacing the adopted maximum FWZI by the 43~GHz $v=1$ measurement changes the principal population fractions only modestly: the AGB fraction with FWZI $<10$~km~s$^{-1}$ changes from 81.5\% to 82.0\%, and the RSG fraction with FWZI $>15$~km~s$^{-1}$ changes from 81.2\% to 78.6\%. Thus, the procedure of adopting the largest available FWZI is not responsible for the overall separation, although intrinsic transition-dependent maser strengths remain relevant for individual sources \citep{2018ApJ...862..153S}.}

{The adopted measurements also do not show a systematic spectral-resolution difference capable of producing the broad RSG distribution. Of the 1,020 AGB stars, 1,000 have adopted FWZI values from the Nobeyama 43~GHz data with an effective velocity resolution of approximately 0.3~km~s$^{-1}$, while 20 have adopted values from the Mopra 86~GHz data with a resolution of approximately 0.2~km~s$^{-1}$. For the 32 RSGs, the adopted measurements comprise 14 Nobeyama spectra (approximately 0.3~km~s$^{-1}$), 13 KVN spectra (0.22~km~s$^{-1}$), three IRAM 30-m spectra (0.14~km~s$^{-1}$), and two ATCA spectra (0.87~km~s$^{-1}$). Thus, 30 of the 32 RSGs were measured at velocity resolutions comparable to or finer than those of the AGB sample. Excluding the two ATCA sources still leaves 24 of 30 RSGs (80.0\%) with FWZI $>15$~km~s$^{-1}$, with a median FWZI of 21.8~km~s$^{-1}$, so the broad RSG distribution is not driven by the two coarser-resolution spectra.}

{Finally, SiO maser variability can change the detectability of weak components. In our dataset, eight AGB stars have repeated observations at different epochs. For this variability check, we compare two observing epochs per source. For the six sources with valid 43~GHz $v=1$ detections at the two epochs compared, the median absolute FWZI change is 3.44~km~s$^{-1}$ (range 0.88--4.80~km~s$^{-1}$); for the eight sources with valid 43~GHz $v=2$ measurements, the corresponding median change is 2.86~km~s$^{-1}$ (range 0.88--6.92~km~s$^{-1}$). These values indicate that variability can add several km~s$^{-1}$ of scatter to an individual FWZI measurement. No comparable multi-epoch RSG sample is available here, however, so a class-dependent variability bias cannot be fully tested. Taken together, the checks above show that observational sensitivity, brightness, transition availability, spectral resolution, and variability can affect individual FWZI measurements, but the tested effects do not remove the population-level difference between the AGB and RSG samples. Accordingly, we use FWZI as an empirical diagnostic and do not interpret the full AGB--RSG difference as a bias-free measure of intrinsic kinematics. Additional diagnostics are therefore required to characterize the remaining ambiguous sources.}

\subsection{Infrared Two-Color Diagrams of AGB Stars and RSGs}
\label{subsec:2_Color_diagram}

The line-width distributions presented in Section~\ref{subsec:FWZI_histogram} provide substantial but incomplete separation between AGB stars and RSGs. We therefore examine whether infrared colors offer a complementary diagnostic.

\begin{figure}
    \centering
    \includegraphics[width=\textwidth]{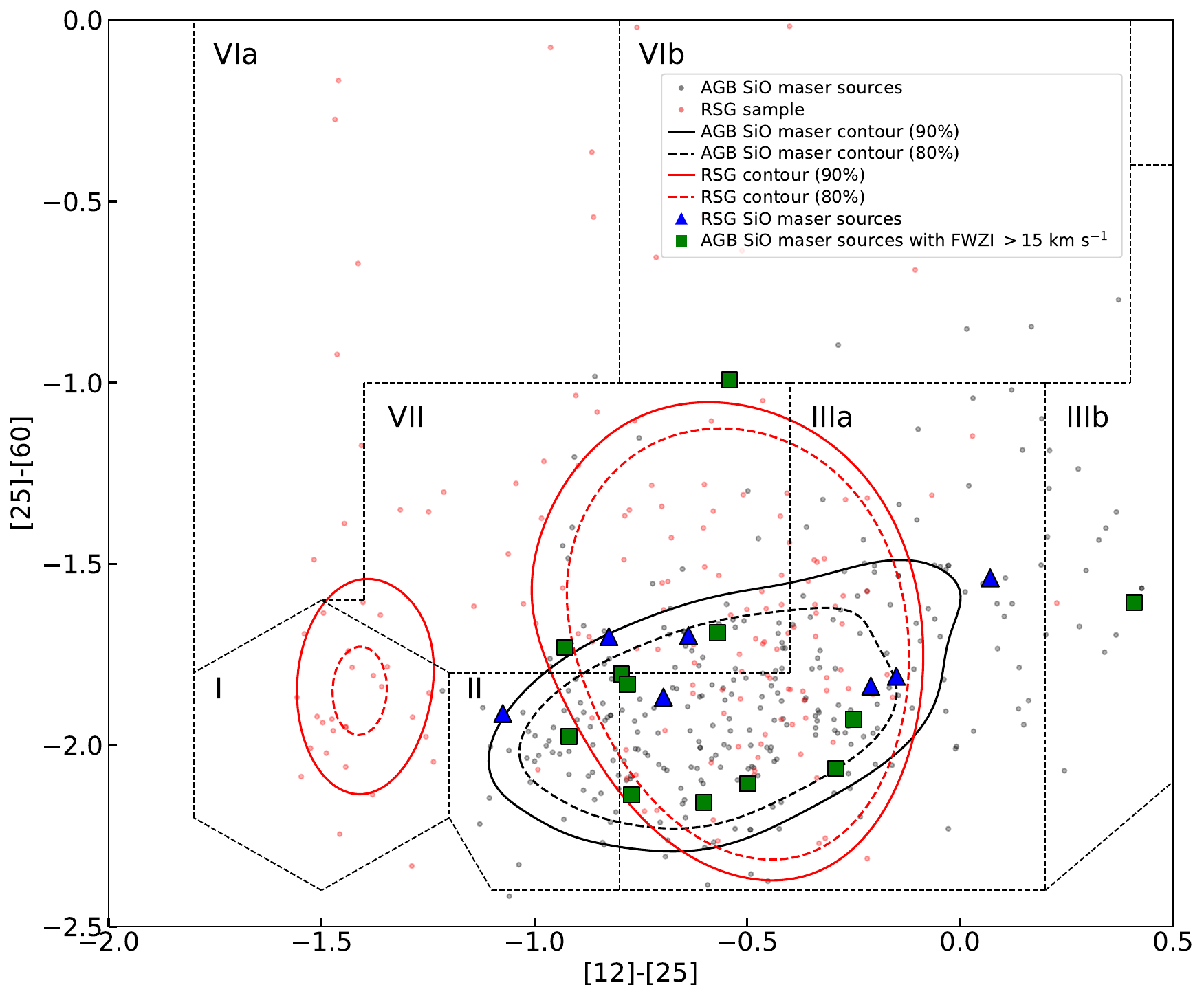}
    \caption{IRAS two-color diagram showing the distributions of the AGB and RSG-candidate populations. The colors are defined as $[12]-[25]=2.5\log_{10}(F_{25}/F_{12})$ and $[25]-[60]=2.5\log_{10}(F_{60}/F_{25})$, where $F_{12}$, $F_{25}$, and $F_{60}$ are the IRAS flux densities at 12, 25, and 60~$\mu$m, respectively. AGB SiO maser sources (gray points; 307 objects) are plotted together with the RSG-candidate comparison sample from \citet{2024MNRAS.529.3630H} (light red points; 178 objects) and the subset of known RSGs with SiO maser detections (blue triangles; 7 objects). To ensure reliable measurements of both colors, only sources with an IRAS flux-quality flag of $\mathrm{FQUAL}=3$ in all three bands were included. AGB SiO maser sources with FWZI $>15$~km~s$^{-1}$ are highlighted as green squares. The solid and dashed contours indicate the 90\% and 80\% regions, respectively, for the AGB and RSG-candidate samples, constructed following the procedure described by \citet{2025ApJ...978..114X}. The dashed boundaries indicate the classical IRAS color regions defined by \citet{1988A&A...194..125V}.}
    \label{fig:IRAS_2color_diagram}
\end{figure}

Figure~\ref{fig:IRAS_2color_diagram} compares the AGB sample with the RSG-candidate comparison sample from \citet{2024MNRAS.529.3630H} in the IRAS two-color diagram. This comparison sample includes both highly probable and likely RSG candidates, and possible contamination by AGB stars remains, especially in the low-luminosity part of the sample. The classical regions defined by \citet{1988A&A...194..125V} are shown for reference. The density distributions of the two populations are offset on average, but their overlap remains substantial.

The overlap is particularly evident among the SiO-maser-emitting RSGs shown by blue triangles, most of which lie within the region densely populated by AGB stars. Several broad-line AGB stars highlighted by green squares also extend into regions occupied by the RSG-candidate comparison sample. Thus, although the IRAS colors distinguish the two populations statistically, they do not cleanly separate SiO-maser-emitting RSGs from broad-line AGB stars.

\begin{figure}
    \centering
    \includegraphics[width=\textwidth]{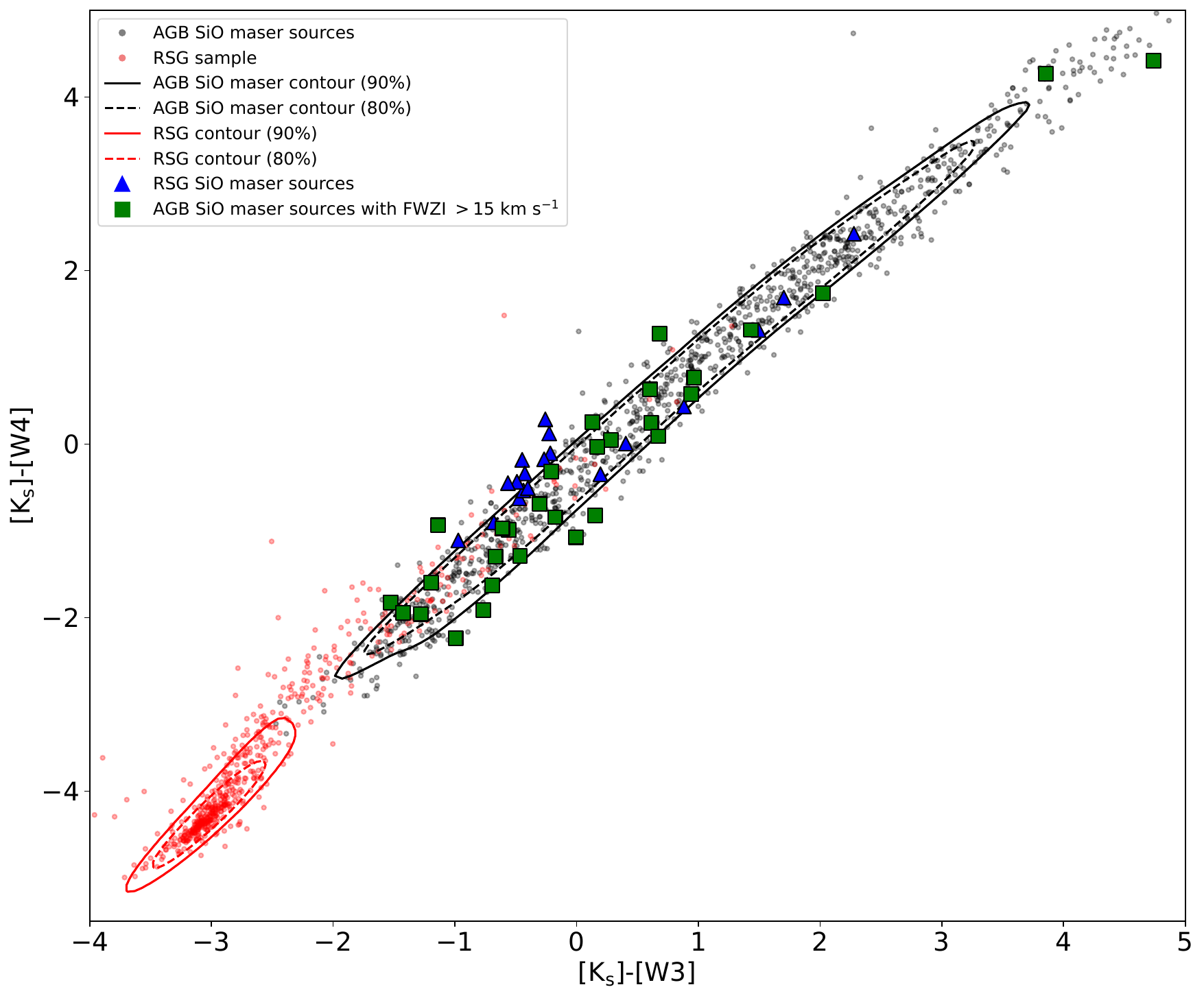}
    \caption{$[K_{\mathrm{s}}]-[W3]$ versus $[K_{\mathrm{s}}]-[W4]$ two-color diagram showing the distributions of the AGB and RSG-candidate populations. AGB SiO maser sources from this study (gray points; 1,061 objects with WISE photometric-quality flags of A or B) are plotted together with the RSG-candidate comparison sample from \citet{2024MNRAS.529.3630H} (light red points; 605 objects with WISE photometric-quality flags of A or B and acceptable 2MASS $K_{\mathrm{s}}$ photometric quality) and the subset of known RSGs with SiO maser detections (blue triangles; 22 objects with WISE photometric-quality flags of A or B). AGB SiO maser sources with FWZI $>15$~km~s$^{-1}$ are highlighted as green squares. The solid and dashed contours indicate the 90\% and 80\% regions, respectively, for the AGB and RSG-candidate samples, constructed in the same manner as in Figure~\ref{fig:IRAS_2color_diagram}.}
    \label{fig:WISE_2MASS_2color_diagram}
\end{figure}

The $[K_{\mathrm{s}}]-[W3]$ versus $[K_{\mathrm{s}}]-[W4]$ diagram in Figure~\ref{fig:WISE_2MASS_2color_diagram} gives a similar result. Because the two colors share the $K_{\mathrm{s}}$ magnitude, part of the pronounced diagonal elongation is inherent to the construction of the diagram. We therefore use this diagram to compare the relative locations and overlap of the samples, rather than interpreting the elongation itself as an independent physical sequence. The overall AGB sample and the RSG-candidate comparison sample occupy different regions on average, but the separation remains incomplete. In particular, the known SiO-maser-emitting RSGs are offset from the density center of the broader RSG-candidate comparison sample and lie closer to the AGB locus.

This offset may indicate that SiO-maser-emitting RSGs represent a particular subset of the broader RSG population, or it may reflect selection effects in the comparison sample. In either case, the two infrared diagrams provide useful population-level information but do not isolate the ambiguous SiO maser sources identified from their line widths. We therefore next combine FWZI with luminosity, a more direct physical parameter for comparing AGB stars and RSGs.

\subsection{SiO Maser Line Width versus Luminosity}
\label{subsec:Final_Candidate}

\begin{figure}
\centering
\includegraphics[width=\textwidth]{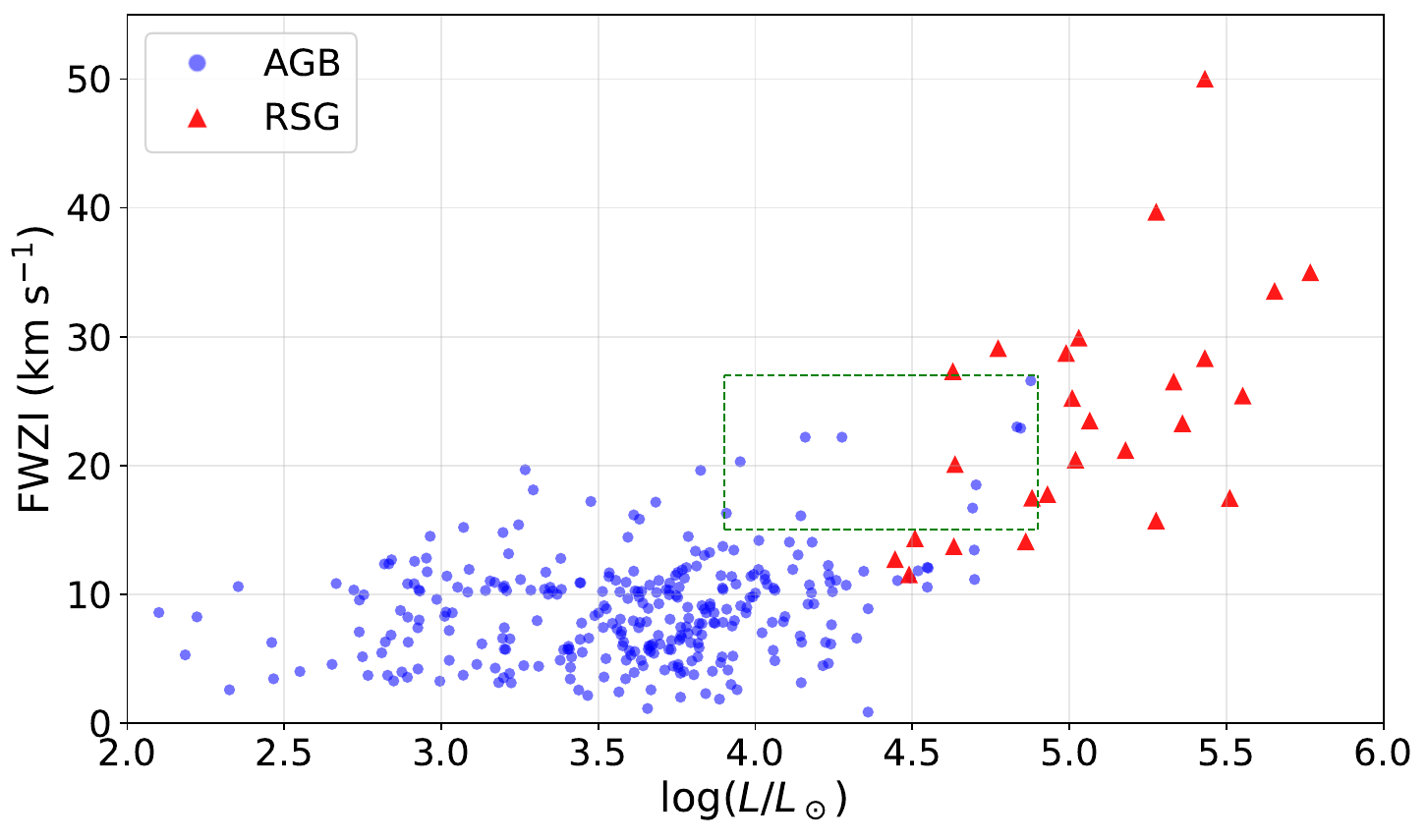}
\caption{{SiO maser line width (FWZI) versus luminosity ($\log(L/L_{\odot})$) for 309 AGB stars and 26 RSGs with adopted distances. The x-axis shows $\log(L/L_{\odot})$, and the y-axis shows FWZI in km~s$^{-1}$. Blue points represent AGB SiO maser sources, and red triangles represent RSG SiO maser sources. Luminosities are based primarily on RUWE-selected Gaia/Bailer-Jones distances; independent literature distances are adopted for five RSGs as described in Section~\ref{subsubsec:RSG_GAIA_RVS}. For visual clarity, luminosity uncertainties are not shown. The effects and limitations of the adopted distances are discussed in Sections~\ref{subsec:Final_Candidate} and~\ref{subsec:Integrated_interpretation}. The green dashed box marks the empirical RSG-adjacent region, defined by $15 \leq \mathrm{FWZI} \leq 26.6$~km~s$^{-1}$ and $3.9 \leq \log(L/L_{\odot}) \leq 4.9$. This region highlights the high-luminosity and broad-line-width tail of the AGB-classified sample, where the sources approach the RSG distribution most closely. The ten AGB-classified objects in this region are listed in Table~\ref{tab:candidates} and are examined further in Sections~\ref{subsec:Advanced_analysis} and~\ref{subsec:Integrated_interpretation}.}}
\label{fig:FWZI_vs_L}
\end{figure}

{Figure~\ref{fig:FWZI_vs_L} shows FWZI as a function of luminosity for 309 AGB stars and 26 RSGs with adopted distances. The luminosities were derived by integrating the infrared SEDs following the procedure described in Section~\ref{subsec:luminosity_estimation}; they are based primarily on RUWE-selected Gaia/Bailer-Jones distances, with the five RSG literature-distance exceptions described in Section~\ref{subsubsec:RSG_GAIA_RVS}.}

The two populations occupy different parts of the diagram on average. AGB stars are concentrated primarily at lower luminosities and narrower SiO maser line widths, whereas RSGs are preferentially found at higher luminosities and broader line widths. The FWZI--luminosity diagram therefore places the empirical line-width difference between the two populations in the context of stellar luminosity.

Nevertheless, the populations overlap substantially. {A weak positive trend is apparent}, with sources having broader SiO maser profiles tending to be more luminous on average, but no clear boundary separates AGB stars from RSGs. In particular, some AGB-classified sources extend into the high-luminosity, broad-line part of the diagram occupied by RSGs. The FWZI--luminosity plane therefore distinguishes the two populations statistically but does not provide an unambiguous classification for every source.

{Because Gaia astrometry can be problematic for large evolved stars, we tested the sensitivity of Figure~\ref{fig:FWZI_vs_L} to the adopted Gaia/Bailer-Jones distances using available VLBI measurements. Among the sources plotted in Figure~\ref{fig:FWZI_vs_L} for which \texttt{R\_med\_geo} is used in the main analysis, VLBI distances are available for PZ~Cas, IRC~$-10414$, and 15060+0947 \citep{2019ApJ...885..131R,2022A&A...667A..74A}. Recalculating the luminosities with these VLBI distances changes $\log(L/L_\odot)$ by only $+0.07$, $+0.20$, and $+0.08$~dex, respectively, and does not change the overall distribution or the membership of the RSG-adjacent region. As an additional check, W~Leo, an AGB star outside the plotted sample with RUWE $<1.4$ and an available VLBI distance \citep{2022A&A...667A..74A}, was included in the distance comparison. For these four validation sources, the ratio of the Bailer-Jones distance \texttt{R\_med\_geo} to the VLBI distance $d_{\rm VLBI}$ ranges from 0.795 to 1.082 and has both positive and negative residuals, providing no indication of a clear systematic offset in this limited sample. The validation sample is small, however, and RUWE $<1.4$ alone does not guarantee an accurate Gaia distance for individual evolved stars \citep{2022A&A...667A..74A}. No VLBI distance was found for any of the ten AGB-classified sources in the RSG-adjacent region; their luminosities therefore remain based on the RUWE-selected Bailer-Jones distances used in the main analysis.}

{We also checked whether the ten AGB-classified sources in the RSG-adjacent region are preferentially nearby, as would be expected if sensitivity effects made broad FWZI values easier to recover only for the closest objects. Their median adopted distance is 2.32~kpc, compared with 2.67~kpc for the remaining 299 AGB stars plotted in Figure~\ref{fig:FWZI_vs_L}. A Mann--Whitney $U$ test gives $U=1368$ and $p=0.649$, providing no evidence that the RSG-adjacent AGB sources are drawn preferentially from smaller distances.}

To characterize the part of the AGB distribution closest to the RSG population, we define the empirical RSG-adjacent region marked by the green dashed box in Figure~\ref{fig:FWZI_vs_L}. It is bounded by $15 \leq \mathrm{FWZI} \leq 26.6$~km~s$^{-1}$ and $3.9 \leq \log(L/L_{\odot}) \leq 4.9$. The lower FWZI boundary and both luminosity boundaries were chosen empirically by visual inspection to encompass the high-luminosity, broad-line tail of the AGB-classified sample lying closest to the RSG distribution. The upper FWZI boundary of 26.6~km~s$^{-1}$ was set by the maximum FWZI of the AGB reference sample. These boundaries are used only as practical criteria for selecting sources for further analysis and have no unique statistical or physical significance.

This region is not intended to define a new stellar class. If some sources within it are genuinely luminous evolved stars with unusually broad SiO maser profiles, they may lie near the high-initial-mass end of the AGB population. Combining FWZI and luminosity may therefore provide a useful means of selecting a small and astrophysically interesting subset of AGB-classified sources for further investigation.

\section{Discussion}\label{sec:discussion}

\subsection{Additional Diagnostics for the AGB-classified Sources in the RSG-adjacent Region}
\label{subsec:Advanced_analysis}

{The FWZI--luminosity selection in Section~\ref{subsec:Final_Candidate} identifies a small number of AGB-classified sources near the region occupied by RSGs, despite substantial overlap between the two populations. Such objects need not all have the same physical origin: some may be relatively low-initial-mass stars observed at an advanced, high-mass-loss stage of AGB evolution, whereas others may belong to the high-initial-mass AGB/SAGB regime. They therefore constitute a potentially useful subset for testing whether FWZI, together with luminosity and other diagnostics, can help distinguish these alternatives observationally.}

The purpose of the following source-by-source analysis is to examine what kinds of objects are found among the AGB-classified sources that occupy the high-luminosity and broad-line-width edge of the AGB distribution. Such sources may be ordinary luminous AGB stars, objects with more RSG-like or SAGB-like properties, or sources whose locations in the FWZI--luminosity plane are affected by uncertainties in distance, photometry, or circumstellar structure.

The ten AGB-classified sources discussed below were selected from the rectangular region marked by the green dashed box in Figure~\ref{fig:FWZI_vs_L}, defined by $15 \leq \mathrm{FWZI} \leq 26.6$~km~s$^{-1}$ and $3.9 \leq \log(L/L_{\odot}) \leq 4.9$. These ten sources are listed in Table~\ref{tab:candidates}.

In this subsection, we examine these sources using three diagnostics beyond the SiO maser line width: (i) infrared SEDs and DUSTY fitting, (ii) OH and H$_2$O maser information, and (iii) period and variability. The information used for this assessment was collected from various archival datasets and catalogs; however, because the availability and quality of these diagnostics are not uniform among the ten sources, the level of constraint differs from object to object. In Section~\ref{subsec:Integrated_interpretation}, we combine the available information and present an integrated interpretation of these sources.

\subsubsection{SED Modeling and DUSTY Fitting} \label{subsec:mass-loss_estimation}

Infrared SEDs of all ten selected sources were modeled using the one-dimensional radiative-transfer code DUSTY \citep{1999astro.ph.10475I}. The purpose of this analysis is not to obtain a unique physical solution for each source, but to assess whether the observed SED can be broadly represented by a simple one-dimensional single-shell model. A satisfactory fit therefore indicates that such a simplified description can reproduce the main SED features, whereas a poor fit indicates that the one-dimensional single-shell approximation is insufficient. Neither outcome by itself uniquely determines the physical geometry or the cause of any discrepancy.

In the DUSTY calculations, the central star was assumed to emit as a blackbody, and the circumstellar dust shell was treated using a one-dimensional spherical-shell approximation. Following the treatment adopted in previous studies \citep{2017MNRAS.465.4482Y,2025arXiv251216209X}, we adopted a 50\%--50\% mixture of ``warm'' and ``cold'' silicate, together with the standard Mathis--Rumpl--Nordsieck (MRN) grain-size distribution \citep{1977ApJ...217..425M}, namely $n(a)\propto a^{-3.5}$, with $a_{\rm min}=0.005\,\mu$m and $a_{\rm max}=0.25\,\mu$m. The radial density distribution followed the analytical prescription of DUSTY for radiation-pressure-driven winds, and the shell thickness was fixed at $R_{\rm out}/R_{\rm in}=10^{4}$.

In the fitting, we varied three parameters: the stellar effective temperature $T_{\rm eff}$, the dust temperature at the inner edge of the dust shell $T_{\rm d}$, and the optical depth at $2.2\,\mu$m, $\tau_{2.2}$. To explore the parameter space, we computed grids spanning $T_{\rm eff}=1400$--$8000$\,K with a step size of 200\,K, $T_{\rm d}=100$--$1500$\,K with a step size of 50\,K, and two optical-depth ranges: $0.01\leq\tau_{2.2}\leq 1$ with a step size of 0.0168, and $1\leq\tau_{2.2}\leq 10$ with a step size of 0.5. For each grid point, the model fluxes were calculated at the observed photometric wavelengths and compared with the observed values after correction for interstellar extinction.

Following \citet{2025arXiv251216209X}, we adopted the infrared extinction law $E(\lambda - V)/E(B - V)$ from \citet{1983MNRAS.203..301H} and assumed a standard total-to-selective extinction ratio of $R_V=3.1$. For each source, the color excess $E(B-V)$ was taken from the Galactic reddening map of \citet{2011ApJ...737..103S} via the NASA/IPAC Infrared Science Archive\footnote{\url{https://irsa.ipac.caltech.edu/applications/DUST/}}, and the corresponding $V$-band extinction was calculated as $A_V=R_V E(B-V)$. The extinction at wavelength $\lambda$ was then derived as
\begin{equation}
A_\lambda = A_V\left[1+\frac{E(\lambda - V)/E(B - V)}{R_V}\right].
\label{eq:extinction}
\end{equation}
The observed fluxes were corrected using
\begin{equation}
F_{\rm corr}(\lambda)=F_{\rm obs}(\lambda)\times10^{0.4A_\lambda}.
\label{eq:flux_correction}
\end{equation}
In practice, this correction was applied only to photometric data at wavelengths shorter than $8\,\mu$m, where interstellar extinction is non-negligible for the present analysis. At longer wavelengths, interstellar extinction was assumed to be negligible, and no correction was applied. This procedure provides a uniform first-order correction for all sources, although for heavily reddened Galactic objects it may overestimate the actual extinction to the source because the reddening map represents the integrated extinction along the line of sight.

As a quantitative criterion for evaluating how well the DUSTY model reproduces the observed SED, we adopted {the same fitting statistic as that used in} \citet{2025arXiv251216209X}. We denote this quantity by $\chi^2_{\rm fit}$:
\begin{equation} \label{eq:chi_fit}
    \chi^2_{\rm fit} = \frac{\sum_i (O_i - E_i)^2/E_i^2}{N_{\rm obs} - N_{\rm para}},
\end{equation}
where $O_i$ is the observed flux in band $i$, and $E_i$ is the flux predicted by the DUSTY model. Here, $N_{\rm obs}$ is the number of observed data points, and $N_{\rm para}$ is the number of grid parameters varied in the fitting procedure, namely $T_{\rm eff}$, $T_{\rm d}$, and $\tau_{2.2}$. Thus, $N_{\rm para}=3$ in the present fitting. For each candidate model, a vertical flux-scaling factor was allowed in order to minimize $\chi^2_{\rm fit}$; this scaling factor was not included in $N_{\rm para}$. Visual inspection was also performed to confirm the overall SED shape. Following \citet{2025arXiv251216209X}, models satisfying $\chi^2_{\rm fit}<9.0\times10^{-4}$ were regarded as acceptable fits.

For each well-fitted object, the total gas+dust mass-loss rate was derived from the best-fit DUSTY model. Following \citet{2025arXiv251216209X}, we adopted \texttt{density type = 4}, which uses the analytic approximation for a radiatively driven wind. In this mode, DUSTY provides a nominal total gas+dust mass-loss rate based on the scaled solution, with tabulated values corresponding to $L=10^4\,L_{\odot}$, $r_{\rm gd}=200$, and $\rho_s=3\,\mathrm{g\,cm^{-3}}$, where $r_{\rm gd}$ is the gas-to-dust mass ratio and $\rho_s$ is the grain material density. The mass-loss rate scales as
\begin{equation}
\dot{M}_{\rm total} \propto L^{3/4}(r_{\rm gd}\rho_s)^{1/2}.
\label{eq:massloss_scaling}
\end{equation}
The nominal DUSTY output for each best-fit model was then rescaled to the luminosity adopted for that source, while $r_{\rm gd}$ and $\rho_s$ were kept fixed at the above values. {Recent observational estimates indicate that the gas-to-dust ratio can vary substantially among evolved stars; for example, \citet{2025A&A...704A.276W} found a median value of $r_{\rm gd}=250$ for O-rich sources, with a broad source-to-source distribution. Adopting $r_{\rm gd}=250$ instead of 200 would increase the DUSTY-derived mass-loss rates by only about 12\% through the $r_{\rm gd}^{1/2}$ scaling and would not affect the SED fits or the qualitative interpretation presented here.} No independent observed expansion velocity was imposed in this rescaling; the mass-loss rates listed in Table~\ref{tab:dustyfits_full} are therefore the total gas+dust mass-loss rates derived from the DUSTY radiatively driven wind model under these assumptions. The DUSTY fitting also provides the effective temperature $T_{\rm eff}$ of the best-fit model, which was used, together with mass-loss properties, variability characteristics, and maser properties, to aid in assessing the likely nature of each source.

A successful DUSTY fit indicates that the observed SED is broadly consistent with a simple one-dimensional single-shell description, but it does not prove that the circumstellar envelope is spherical. Small-scale asymmetric structures, such as geometrically thick tori with only modest deviations from a roughly spherical shell or recently formed compact bipolar outflows, may not be captured by this modeling. Conversely, a poor fit may reflect multiple dust components, uncertainty in the extinction correction, limitations of the available photometry, counterpart-identification problems, or more complex circumstellar geometry. Uncertainties in the interstellar-extinction correction may also affect the interpretation of whether the observed SED can be represented by such a simple one-dimensional model \citep{2025arXiv251216209X}. The DUSTY fitting results should therefore be used as supplementary constraints rather than as definitive evidence for the circumstellar geometry.

For the sources that are well fitted by the single-shell models (Figure~\ref{fig:dusty_fit_sed}), the derived parameters, including effective temperature, dust temperature, optical depth, and mass-loss rate, are used as supplementary constraints in the subsequent discussion. Here, ``well fitted'' means that the best-fit model satisfies $\chi^2_{\rm fit}<9.0\times10^{-4}$ and that visual inspection shows no severe deviations in the main near- to mid-infrared SED shape. The fitting results are summarized in Table~\ref{tab:dustyfits_full}.

By contrast, two sources show poor fits (Figure~\ref{fig:fail_sed}). In these cases, the model does not reproduce the observed SED shape well. Such poor fits indicate that the simple one-dimensional single-shell approximation is inadequate for these sources, but they do not provide direct evidence for a specific physical cause. In the integrated interpretation of Section~\ref{subsec:Integrated_interpretation}, the well-fitted sources therefore provide DUSTY-derived parameters that can be used as supporting constraints, whereas the poorly fitted sources remain less well constrained and must be interpreted more cautiously.

{For additional context, MSX SMC~055 (IRAS~00483$-$7347), a well-studied observational SAGB candidate in the SMC \citep{2009A&A...506.1277G,2017PASA...34...56D}, is included in Table~\ref{tab:dustyfits_full} as a comparison object, and its SED and DUSTY fit are shown in Figure~\ref{fig:MSX_SMC_055}. The motivation and limitations of this comparison are discussed in Section~\ref{subsec:Integrated_interpretation}.}

\subsubsection{Maser Information}

This subsection focuses on maser information as a circumstellar diagnostic. In addition to the SiO maser properties that define the sample, we also consider the available H$_2$O and OH maser information, as summarized source by source in Appendix~\ref{app:individual_notes}. Together, these maser data provide complementary constraints on the circumstellar environment and help assess whether a source is more consistent with a relatively compact inner-envelope structure or with a more extended and developed circumstellar envelope \citep[e.g.,][]{1996A&ARv...7...97H,2012JKAS...45..139K}.

{In this subsection, H$_2$O maser emission refers to the $6_{16}$--$5_{23}$ transition at 22.235~GHz. For OH, we distinguish the 1612~MHz satellite line from the 1665 and 1667~MHz main lines \citep[e.g.,][]{2010ApJS..188..209K,2013AJ....145...22K,2025ApJ...978..114X}.}

Different maser species probe different regions of the circumstellar envelope. SiO maser emission is generally associated with dense gas in the inner envelope close to the stellar atmosphere, whereas H$_2$O maser emission typically traces active outflow and molecular gas at somewhat larger radii, and OH maser emission often indicates a more extended outer envelope and sustained mass loss \citep[e.g.,][]{2007A&A...470..191W,2012JKAS...45..139K}. Therefore, for sources in which multiple maser species have been reported, line-profile information and, where available, interferometric maps can provide structural and kinematic information over different regions of the circumstellar envelope.

{A classical framework for interpreting the coexistence of circumstellar masers is the chronological sequence proposed by \citet{1989ApJ...338..234L}, in which SiO, H$_2$O, and OH masers appear and disappear as the circumstellar envelope develops. This sequence is useful as a population-level guide, but it should not be treated as a strict evolutionary clock for individual stars. Because the H$_2$O and especially OH masing zones lie farther from the star than the SiO maser region, their presence or absence can be affected by the structure of the outer envelope; non-spherical structure and binary interaction can further modify circumstellar morphology and kinematics \citep[e.g.,][]{2020Sci...369.1497D}. We therefore use the multi-species maser pattern only as supplementary evidence rather than as a unique indicator of evolutionary state.}

When available, maser profile properties are also informative. The profile width or full velocity extent, the presence of multiple peaks, asymmetry, and overall velocity spread can all reflect the kinematics of the emitting gas. For example, broader or more structured profiles may indicate a larger range of line-of-sight velocities, multiple masing clumps, or complex outflow geometry, whereas a narrow or simple profile may suggest a more limited velocity field. For OH and H$_2$O masers in particular, such profile information can provide useful supporting evidence for expansion, acceleration, or asymmetry in the circumstellar envelope \citep[e.g.,][]{2003ApJ...599.1372D,2024AJ....168...53Y,2025ApJ...978..114X}.

{The OH transitions are not interchangeable diagnostics. The relative detectability of the 1612~MHz satellite line and the 1665/1667~MHz main lines depends on circumstellar-shell properties; in relatively thin shells the 1612~MHz line can be weak, while main-line emission may still be detected \citep[e.g.,][]{1995A&AS..111..237L,1996A&ARv...7...97H}. In addition, OH main-line emission can be strongly variable, and main-line velocity components need not coincide with the characteristic double-peaked 1612~MHz profile \citep[e.g.,][]{2025ApJ...978..114X}. We therefore interpret OH profile morphology and velocity extent with the relevant transition explicitly in mind.}

At the same time, a non-detection of a maser species, when the source has been observed, can also provide some information on the evolutionary stage, local physical conditions, or variability of the source. However, the diagnostic value of a non-detection should be treated with caution. Non-detection does not necessarily imply the absence of the maser species, because survey sensitivity, frequency coverage, sky coverage, and epoch-to-epoch variability differ substantially among catalogs and observations \citep[e.g.,][]{2012JKAS...45..139K,2024ApJS..275...20L}. For this reason, maser information should be interpreted as supplementary evidence rather than as a standalone classification criterion.

Appendix~\ref{app:individual_notes} provides source-by-source notes that compile the available SiO, H$_2$O, and OH maser information for each object, together with supplementary details on profile properties, velocity extent, and literature-based detections and non-detections. These notes document the maser evidence in a more granular form than is practical in the main text, and they are used as part of the basis for the integrated interpretation in Section~\ref{subsec:Integrated_interpretation}. In the source-by-source assessment below, we therefore use the available SiO, H$_2$O, and OH maser information, together with the notes compiled in Appendix~\ref{app:individual_notes}, to judge whether each source is more consistent with an AGB-like object, a more RSG-like object, or a source that remains difficult to classify.

\subsubsection{Period and Variability}

Period information is used here only as supporting evidence and is interpreted together with the other diagnostics in the next subsection. For evolved cool stars, long-term variability is often related to radial pulsation and circumstellar activity, so a period measurement can provide useful context for assessing the nature of a source, even though it does not by itself distinguish AGB stars from RSGs \citep[e.g.,][]{2009asrp.proc...48F,2023A&A...674A..15L}. In the integrated discussion below, the period information is therefore used as supplementary information to help place each source into one of three qualitative categories: AGB-like, more RSG-like, or difficult to classify.

{We examined the NEOWISE W2 light curves for all ten sources,} using the Lomb--Scargle periodogram to search for periodicity \citep{1976Ap&SS..39..447L,1982ApJ...263..835S}, following the procedure adopted by \citet{2025ApJ...978..114X}. The WISE and NEOWISE data provide mid-infrared time-domain photometry that is useful for examining long-term variability in red evolved stars \citep{2010AJ....140.1868W,2014ApJ...792...30M}. Two sources, J17572314$-$1802456 and J18420731$-$0423334, have usable NEOWISE W2 data, but their light curves exhibit irregular variability without a significant periodic signal. The remaining sources do not have usable NEOWISE W2 photometry for period analysis.

The lack of detectable periodicity in the two sources should be interpreted with caution. The available light curves are affected by a limited number of epochs, uneven sampling, and seasonal gaps, which can introduce window-function effects and aliasing. Therefore, the absence of a significant period does not necessarily imply true non-variability. Accordingly, the NEOWISE entry ``Non-periodic'' in Table~\ref{tab:candidates} denotes sources with usable NEOWISE W2 data but no significant periodicity, while ``no data'' indicates that no usable NEOWISE W2 light curve is available.

\begin{figure}
\centering
\includegraphics[width=\textwidth]{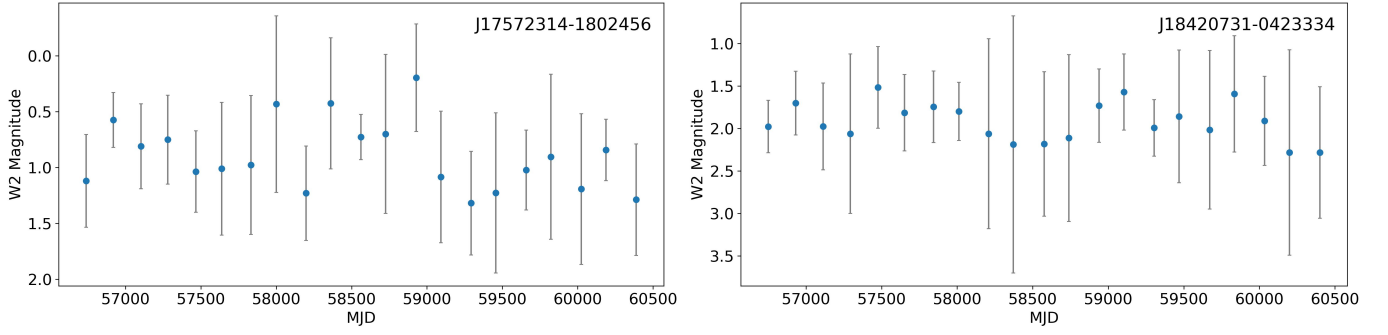}
\caption{NEOWISE W2 light curves for the two sources with usable NEOWISE W2 photometry. Only measurements with signal-to-noise ratios greater than 20, an image-quality score greater than 0, and a frame-quality flag greater than 0 were retained. These two objects are shown because they are the only sources in the selected sample with sufficiently reliable NEOWISE W2 photometry for constructing binned light curves. Time binning was applied using 14-day windows in order to suppress short-timescale scatter and make the long-term variability behavior easier to discern. The blue points represent the mean magnitude within each bin, where multiple observations obtained within each 14-day window were combined into a single epoch. The vertical error bars indicate the within-bin scatter. For bins containing only a single valid measurement, a representative uncertainty equal to the median within-bin scatter of the multi-measurement bins of the same source was assigned. These light curves are suitable for a qualitative variability check, but they do not support a secure period detection.}
\label{fig:WISE_Lightcurve}
\end{figure}

Figure~\ref{fig:WISE_Lightcurve} illustrates the observed variability behavior of these two sources. For this figure, the NEOWISE W2 measurements were binned into 14-day windows in order to suppress short-timescale scatter and make the long-term variability behavior easier to discern. The plotted points represent the mean magnitude in each bin, and the error bars indicate the within-bin scatter. For bins containing only a single valid measurement, we assign a representative uncertainty equal to the median within-bin scatter of the multi-measurement bins of the same source. Although these binned light curves are useful for a qualitative variability check, they do not support a secure period detection for either source.

For the remaining eight sources, period estimates are adopted from external catalogs. Gaia DR3 provides periods for three sources, while the AAVSO International Variable Star Index provides periods for seven sources, with two sources appearing in both catalogs \citep{2023A&A...674A..15L,2006SASS...25...47W}. These catalog periods are useful as supplementary information because they indicate whether the source shows long-term variability consistent with an evolved pulsating star. At the same time, these periods should still be regarded as supporting evidence only, since catalog values may be based on different wavelength bands, different time baselines, and different sampling cadences or data-quality criteria. The period and variability information is summarized in Table~\ref{tab:candidates}. In the next subsection, these constraints are combined with line width, luminosity, SED fitting, and maser information to assess whether each source is more naturally grouped as AGB-like, more RSG-like, or difficult to classify.

\subsection{Integrated Interpretation of the RSG-adjacent Sources}\label{subsec:Integrated_interpretation}

{The ten sources selected for further analysis occupy the RSG-adjacent region in the FWZI--luminosity plane shown in Figure~\ref{fig:FWZI_vs_L}. This region is not meant to define a new stellar class. Rather, it is an empirical parameter space where AGB-classified SiO maser sources approach the luminosity and line-width regime occupied by RSGs. Throughout this subsection, ``AGB-classified'' refers to the catalog-based classification adopted in this paper, not to an independent determination of either the detailed AGB evolutionary phase or the initial-mass regime of each source. Conversely, the RSG comparison sources are identified from external RSG classifications. Therefore, overlap in the FWZI--luminosity plane means similarity in these two empirical observables, but does not by itself imply that the sources have the same physical nature. The central question is whether the selected AGB-classified sources can be explained as relatively low-initial-mass stars in a late, high-mass-loss AGB phase, or whether some instead show evidence favoring a higher-initial-mass AGB/SAGB-related interpretation.}

{This question is astrophysically important because SAGB stars occupy an initial-mass regime near the upper mass boundary for stars that experience AGB-like thermally pulsing evolution; the relevant mass boundaries depend on metallicity and stellar-model physics \citep{2017PASA...34...56D}. This is conceptually different from identifying a star as ``late AGB,'' which describes how far it has progressed through the AGB phase and does not by itself imply a high initial mass. Observational candidates in the SAGB regime are rare and difficult to identify because the expected properties overlap strongly with those of massive AGB stars and RSGs. Therefore, even if the present data cannot confirm any source as a SAGB star, the RSG-adjacent region is potentially useful if it can isolate a small number of AGB-classified sources whose luminosities, maser kinematics, variability, and circumstellar properties make them promising targets for further investigation.}

{In this context, we distinguish two physically different axes that can otherwise be conflated. A late-AGB interpretation concerns evolutionary phase: relatively low-initial-mass stars, including those with initial masses of order a solar mass, can reach an advanced thermally pulsing AGB stage with strong mass loss and a thick circumstellar envelope. A high-mass AGB or SAGB-related interpretation instead concerns primarily the initial-mass regime, with the SAGB boundaries also depending on metallicity. No single observable used here provides a clean separation between these possibilities. Thick dust shells and strong mass loss can occur in both, while high luminosity and long pulsation period can favor a higher initial mass but are not unique identifiers. Population information such as concentration toward the Galactic plane can also be informative \citep{1994MNRAS.267..711W}. Our contribution is to add inner-envelope SiO maser velocity extent to this set of diagnostics: a broad FWZI combined with high luminosity and the other indicators can flag sources for which a high-initial-mass interpretation deserves particular attention, but FWZI is not itself a direct mass measurement. We therefore use SAGB-like or SAGB-related terminology only as a tentative follow-up context, not as a determination of evolutionary phase or initial-mass category.}

\begin{figure}
\centering
\includegraphics[width=0.90\textwidth]{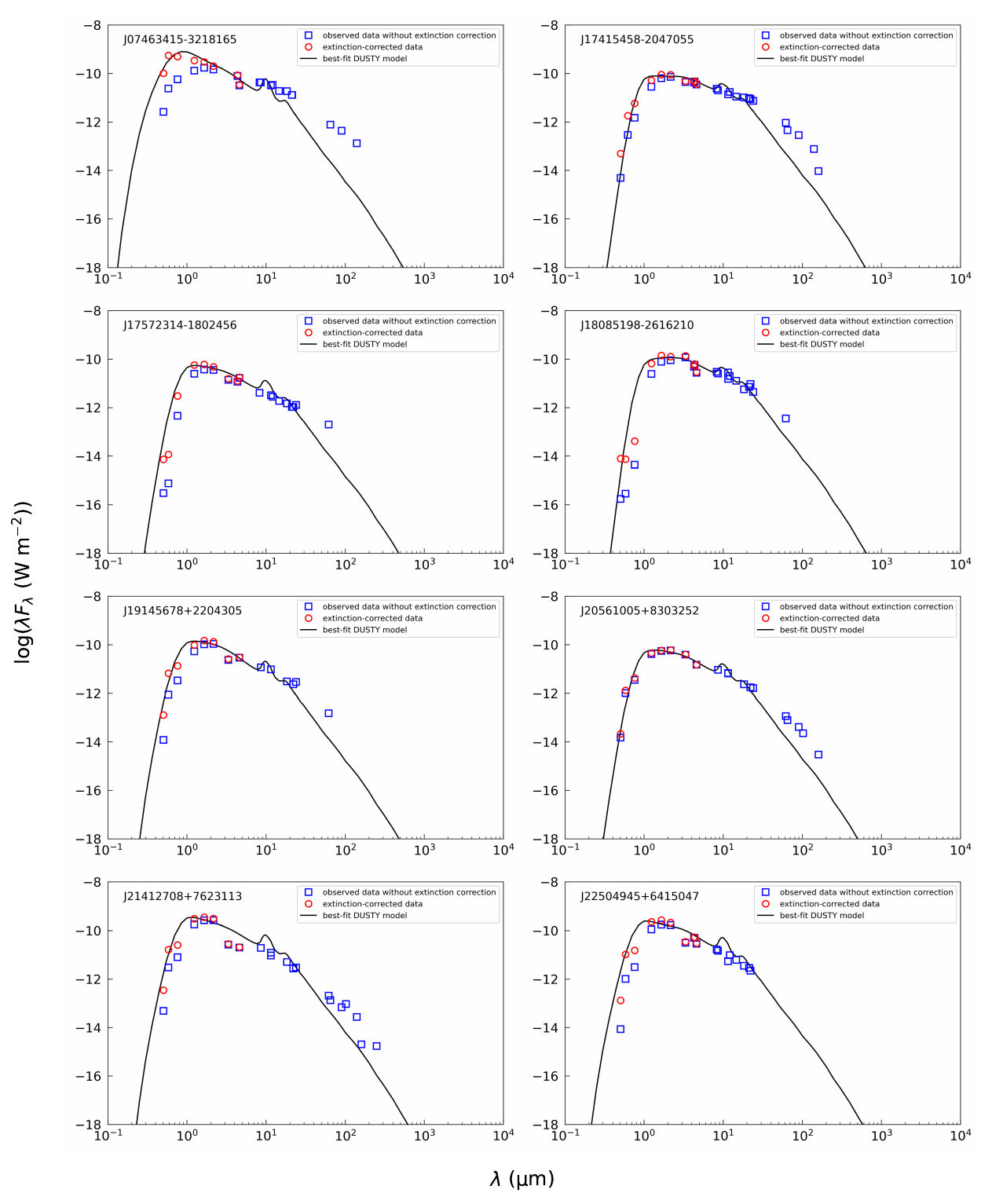}
\caption{Spectral energy distributions (SEDs) of the eight well-fitted sources among the ten selected sources listed in Table~\ref{tab:candidates}. The SED peaks and the overall near- to mid-infrared shapes of these sources are broadly reproduced by the single-shell DUSTY models, although some deviations remain at longer wavelengths. The blue squares show the observed data points without interstellar extinction correction, while the red circles indicate the data points corrected for extinction using the law of \citet{1983MNRAS.203..301H}, applied only to wavelengths shorter than 8~$\mu$m. The black lines represent the best-fit DUSTY models. Here, the agreement refers to the ability of a simple one-dimensional single-shell model to reproduce the main SED features and does not imply a unique physical geometry. The corresponding DUSTY-derived parameters are summarized in Table~\ref{tab:dustyfits_full}. Photometric error bars are not shown for clarity. The two sources that cannot be satisfactorily reproduced by a simple single-shell DUSTY model are presented separately in Figure~\ref{fig:fail_sed}.}
\label{fig:dusty_fit_sed}
\end{figure}

\begin{figure}
\centering
\includegraphics[width=\textwidth]{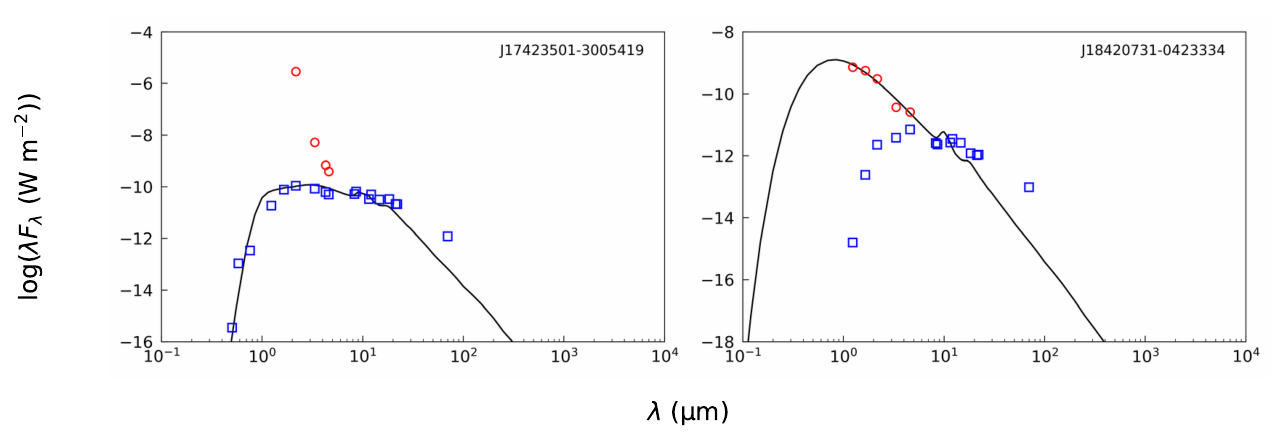}
\caption{SEDs of the two sources with poor DUSTY fits. The blue squares show the observed data points without interstellar extinction correction, while the red circles indicate the data points corrected for extinction using the law of \citet{1983MNRAS.203..301H}, applied only to wavelengths shorter than 8~$\mu$m. The black lines represent the best-fit models obtained with the single-shell DUSTY framework. Photometric error bars are not shown for clarity. Unlike the eight sources shown in Figure~\ref{fig:dusty_fit_sed}, whose SEDs are broadly reproduced by a simple single-shell DUSTY model, these two sources cannot be satisfactorily reproduced by such a model. These poor fits indicate that the simple one-dimensional single-shell approximation is inadequate for these sources, but they do not by themselves identify a unique physical cause. Consequently, the nature and classification of these sources remain uncertain.}
\label{fig:fail_sed}
\end{figure}

{The interpretation below is necessarily tentative. The basic properties are summarized in Table~\ref{tab:candidates}, the DUSTY fitting results are given in Table~\ref{tab:dustyfits_full} and Figures~\ref{fig:dusty_fit_sed} and~\ref{fig:fail_sed}, and the source-by-source maser information, including the relevant OH/H$_2$O detections and the cases with no available information, is compiled in Appendix~\ref{app:individual_notes}. Three caveats are important. First, the luminosities in Figure~\ref{fig:FWZI_vs_L} depend on adopted distances and infrared SEDs. Most distances are RUWE-selected Gaia/Bailer-Jones estimates, although independent literature distances are used for five RSGs. The VLBI comparison in Section~\ref{subsec:Final_Candidate} reveals no clear systematic offset in the small validation sample, but this does not establish the reliability of every Gaia distance; in particular, RUWE $<1.4$ is not by itself sufficient to guarantee a reliable distance for individual AGB stars \citep{2022A&A...667A..74A}. Possible mismatches among the SiO maser position, infrared counterpart, and Gaia source also cannot be completely excluded, especially in crowded Galactic-plane fields. Second, a broad SiO FWZI is not a direct measurement of terminal velocity or stellar mass, but only indicates a large velocity extent of the detected inner-envelope SiO maser emission. Third, the selected sources are drawn from a SiO-maser-selected sample affected by infrared color selection, catalog cross-matching, and Gaia counterpart availability.}

{Binarity provides an additional possible explanation for unusually broad or complex SiO velocity extents. ALMA observations of AGB winds have revealed widespread non-spherical structures that can be explained by interaction with stellar or substellar companions \citep{2020Sci...369.1497D}. However, this and many related studies primarily trace thermal molecular emission, including ground-state ($v=0$) lines, whereas the FWZI measurements used here are based on vibrationally excited ($v>0$) SiO maser emission. It therefore remains unclear whether companion-driven circumstellar kinematics necessarily produce anomalously broad $v>0$ SiO maser profiles. High-angular-resolution and multi-epoch $v>0$ SiO maser observations, ideally combined with thermal molecular-line imaging and independent constraints on binarity, would provide a useful follow-up test.}

With these limitations in mind, the ten sources can be divided into three groups. The first group consists of luminous AGB-like sources: objects that enter the RSG-adjacent region because they are moderately broad-lined or relatively luminous, but whose other diagnostics can still be understood within the range of ordinary or luminous AGB behavior. The second group consists of sources for which a high-initial-mass AGB, RSG-like, or SAGB-related follow-up context is most strongly motivated in the present sample: objects that combine high luminosity, broad SiO FWZI, long pulsation period, and outer-envelope maser activity. This second group is not defined by a single diagnostic that separates SAGB stars from luminous AGB stars; rather, it is defined by {the coincidence of several properties that are not fully accounted for by a simple luminous-AGB interpretation and that make follow-up observations especially valuable}. The third group consists of sources whose SEDs are not reliably reproduced by the simple single-shell DUSTY model, so that the SED/DUSTY constraints are less useful for interpreting their physical nature. The first and second groups together correspond to the eight sources whose SEDs and DUSTY fits are shown in Figure~\ref{fig:dusty_fit_sed}, whereas the third group corresponds to the two sources whose poor SED fits are shown in Figure~\ref{fig:fail_sed}.

\subsubsection{{Group 1: Luminous AGB-like Sources}}

The first group includes J17415458$-$2047055, J17572314$-$1802456, J19145678+2204305, J20561005+8303252, J21412708+7623113, and J22504945+6415047. These sources are {not entirely typical AGB stars} in the sense that several of them have individual properties that place them near the boundary region. For example, J20561005+8303252 has a relatively broad FWZI of 22.2~km~s$^{-1}$, but its luminosity is only $\log(L/L_\odot)=4.159$ and no OH or H$_2$O maser information is available in the literature compiled in Appendix~\ref{app:individual_notes}. J22504945+6415047 is relatively luminous, with $\log(L/L_\odot)=4.692$, but its FWZI is only 16.7~km~s$^{-1}$ and no OH or H$_2$O maser information is available. J21412708+7623113 has a reported H$_2$O maser detection \citep{2013AJ....145...22K}, but its luminosity is relatively low, $\log(L/L_\odot)=3.952$, and its AAVSO period is 333~days. Thus, the boundary-like properties in this group occur individually or only weakly in combination. These sources do not show the combined pattern of high luminosity, long-period variability, enhanced mass-loss activity, and outer-envelope maser activity that would make a high-initial-mass AGB, RSG-like, or SAGB-related interpretation necessary. Their SEDs are also broadly reproduced by a simple single-shell DUSTY model, consistent with a relatively simple dusty AGB-like envelope. Although such fits do not prove spherical geometry, they support the view that these six sources are most naturally interpreted at present as luminous AGB-like objects.

\subsubsection{{Group 2: Sources Motivating High-initial-mass AGB/RSG-like/SAGB-related Follow-up}}

{The second group consists of J07463415$-$3218165 and J18085198$-$2616210. These are the most relevant sources in the present sample for testing whether the RSG-adjacent region contains objects in the high-initial-mass AGB/SAGB regime rather than only relatively low-initial-mass stars in a late AGB phase.} J07463415$-$3218165 has a broad FWZI of 26.6~km~s$^{-1}$, a high luminosity of $\log(L/L_\odot)=4.877$, a long pulsation period of up to 844.5~days, and both OH and H$_2$O maser detections; the OH and H$_2$O information is summarized in Appendix~\ref{app:individual_notes} and is based on \citet{1995A&AS..111..237L} and \citet{2010ApJS..188..209K}. J18085198$-$2616210 has a broad FWZI of 22.9~km~s$^{-1}$, a high luminosity of $\log(L/L_\odot)=4.845$, a {720-day} period, and an OH maser detection reported by \citet{1991A&AS...90..327T}; no H$_2$O maser information is currently available. These objects may still be extreme luminous AGB stars, and the present data are insufficient to classify them as RSGs or SAGB stars. Nevertheless, their combination of broad inner-envelope SiO maser kinematics, high luminosity, long-period variability, and outer-envelope maser activity makes them physically distinct from the first group. They are therefore the two sources in the selected sample for which follow-up in the high-initial-mass AGB, RSG-like, or SAGB-related context is most strongly motivated. This should not be interpreted as evidence that they are confirmed SAGB stars or even high-confidence SAGB candidates.

{Because one motivation for examining the RSG-adjacent region is the possible presence of objects in the high-initial-mass AGB/SAGB regime, we compared J07463415$-$3218165 and J18085198$-$2616210 with MSX SMC~055, a previously proposed SAGB candidate \citep{2017PASA...34...56D}.} {MSX SMC~055 (IRAS~00483$-$7347) was identified by \citet{2009A&A...506.1277G} as a good SAGB candidate on the basis of its very long pulsation period of 1749~days and high luminosity, $M_{\rm bol}\simeq -8.0$, and was subsequently highlighted by \citet{2017PASA...34...56D} as a particularly strong observational candidate. Its inclusion here is therefore motivated primarily by its proposed high-initial-mass AGB/SAGB nature, rather than simply by its location in the low-metallicity SMC environment.}

\begin{figure}
\centering
\includegraphics[width=0.5\textwidth]{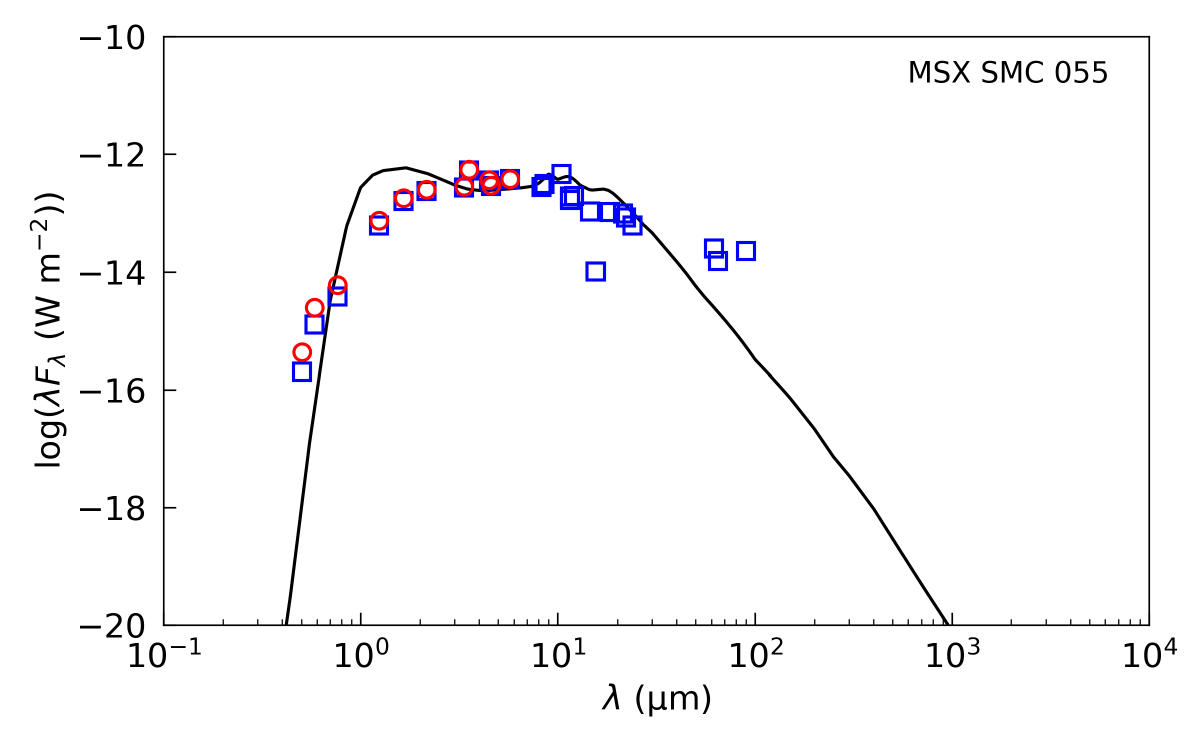}
\caption{{SED of MSX SMC 055 (IRAS~00483$-$7347), an observational SAGB candidate identified on the basis of its unusually long pulsation period and high luminosity \citep{2009A&A...506.1277G,2017PASA...34...56D}, shown here as a contextual comparison object. The blue squares represent the observed photometric data without interstellar extinction correction, while the red circles indicate the extinction-corrected data. The black line shows the best-fit single-shell DUSTY model obtained in this study. The plotting format follows that adopted in Figure~\ref{fig:dusty_fit_sed}, and the corresponding DUSTY-derived parameters are summarized in Table~\ref{tab:dustyfits_full}. Because MSX SMC 055 belongs to the lower-metallicity SMC environment, its DUSTY-derived mass-loss rate is not used as a direct quantitative discriminator relative to the Galactic sources. Its inclusion here does not imply that the selected Galactic sources have the same physical nature as MSX SMC 055.}}
\label{fig:MSX_SMC_055}
\end{figure}

{Figure~\ref{fig:MSX_SMC_055} shows the SED of MSX SMC~055 as a contextual reference object. This comparison is intended to test whether the two second-group sources resemble a previously discussed SAGB candidate in their SED shape and selected DUSTY-derived properties. No strong overall similarity is evident. As listed in Table~\ref{tab:dustyfits_full}, MSX SMC~055 has $\tau_{2.2}=0.983$, $T_{\rm d}=550$~K, and a DUSTY-derived $\dot{M}_{\rm total}=3.0\times10^{-5}\,M_\odot\,{\rm yr}^{-1}$ under the common assumptions adopted in this study. By comparison, J07463415$-$3218165 has a substantially lower optical depth ($\tau_{2.2}=0.161$), a higher inner dust temperature ($T_{\rm d}=1000$~K), a DUSTY-derived $\dot{M}_{\rm total}=5.0\times10^{-6}\,M_\odot\,{\rm yr}^{-1}$, and a much shorter period. J18085198$-$2616210 is closer to MSX SMC~055 in optical depth ($\tau_{2.2}=0.782$) and has a DUSTY-derived $\dot{M}_{\rm total}=1.3\times10^{-5}\,M_\odot\,{\rm yr}^{-1}$, but it likewise has a higher inner dust temperature ($T_{\rm d}=900$~K) and a much shorter period.}

{The metallicity difference must be kept in mind when interpreting the mass-loss comparison. MSX SMC~055 belongs to the SMC environment, where dust formation, the gas-to-dust ratio, and circumstellar-wind properties can differ from those of Galactic evolved stars; for example, \citet{2018MNRAS.473.3835G} modeled SMC sources using a gas-to-dust ratio of 1000 and a metallicity of one-fifth solar. We therefore retain the uniformly derived DUSTY mass-loss rate for MSX SMC~055 in Table~\ref{tab:dustyfits_full} for reference, but do not use its numerical value as a direct quantitative discriminator between MSX SMC~055 and the Galactic sources. Figure~\ref{fig:MSX_SMC_055} should instead be regarded as a qualitative, contextual comparison. It does not demonstrate that either Galactic source has the same physical nature as MSX SMC~055 or that either is a confirmed SAGB star.}

\subsubsection{{Group 3: Sources with Poorly Constrained SEDs}}

The third group consists of J17423501$-$3005419 and J18420731$-$0423334. Their defining feature is that their SEDs are not satisfactorily reproduced by the simple single-shell DUSTY model. In this group, the poor fits do not by themselves define an evolutionary class; rather, they indicate that the SED-based constraints are less reliable or that the circumstellar environments may be more complex than assumed in the model. The two sources should therefore be considered separately. J17423501$-$3005419 is highly luminous, with $\log(L/L_\odot)=4.833$, has a broad FWZI of 23.0~km~s$^{-1}$, a long period of 760.1~days, and both OH and H$_2$O maser detections. The OH and H$_2$O maser information is summarized in Appendix~\ref{app:individual_notes} and is based on \citet{1995A&AS..111..237L} and \citet{2010ApJS..188..209K}. Its poor SED fit may reflect multiple dust components, uncertain extinction correction, counterpart-identification problems, or a genuinely non-spherical circumstellar structure. In particular, the combination of broad SiO emission and OH/H$_2$O maser activity may point to kinematics more complex than a simple spherical AGB wind, such as asymmetric or bipolar mass loss, as discussed in related maser studies \citep[e.g.,][]{2025arXiv251216209X}. Thus, J17423501$-$3005419 is not merely a poorly constrained source, but a broad-line, high-luminosity, maser-active object that deserves special follow-up. By contrast, J18420731$-$0423334 is {less well constrained} because it has no reliable catalog period and no available OH/H$_2$O maser information in the literature compiled in Appendix~\ref{app:individual_notes}. Although usable NEOWISE W2 data are available, Figure~\ref{fig:WISE_Lightcurve} does not support a secure period detection.

{The RSG-adjacent region should therefore be viewed not as a clean classification box for a specific type of object, but as a physically mixed parameter space in which several kinds of evolved objects can coexist. Many sources in this region are likely luminous AGB-like objects, including stars whose circumstellar envelopes have become prominent late in AGB evolution, but some may instead belong to a higher-initial-mass AGB/SAGB regime, and others may have complex circumstellar structures or uncertain counterpart identifications. This mixture is precisely what makes the region interesting: the FWZI--luminosity selection may help separate, for follow-up purposes, stars whose extreme appearance is primarily a consequence of advanced AGB evolution from those for which a high initial mass is a plausible contributing factor. No individual source, however, can be assigned to the SAGB regime from the present data alone.}

A broader implication of this result is that the FWZI--luminosity plane may become a useful screening tool, rather than a definitive classifier, for identifying unusual evolved stars. To apply this method beyond the small sample studied here, large and well-curated SiO maser spectral databases, or catalogs in which FWZI values are measured in a uniform and documented way, are essential. Cataloged peak velocities or detection flags alone are not sufficient for this purpose. Since many SiO maser surveys have been conducted after the Nobeyama surveys and have reported additional detections, extending this approach to larger samples is possible in principle, provided that comparable FWZI measurements can be obtained. When such spectral or uniformly measured FWZI information is combined with infrared photometry and Gaia counterpart information, the RSG-adjacent region can be searched systematically for high-priority targets such as luminous broad-line AGB stars, possible RSG-like or SAGB-related objects, and sources with complex circumstellar envelopes.

\section{Summary}\label{sec:summary}

In this work, we investigated the relationship between SiO maser line-profile properties and the nature of their evolved-star host sources. In particular, we examined whether the SiO maser line width, together with luminosity and additional circumstellar diagnostics, can be used to distinguish AGB stars from RSGs and to identify astrophysically interesting AGB-classified objects in the RSG-adjacent region of the FWZI--luminosity plane. Our main results are summarized as follows:

\begin{enumerate}

\item We compiled a large SiO maser spectral dataset consisting of 2,956 spectra detected from 1,551 objects. {Using this dataset and additional literature data,} we constructed reference samples of 1,068 O-rich AGB stars and 36 RSGs with detected SiO maser emission. {The full width at zero intensity (FWZI) could be reliably measured} for 1,020 AGB stars and 32 RSGs, and these subsamples were used to compare the line-width distributions of the two populations. The two populations show systematically different FWZI distributions: AGB stars are strongly concentrated at narrow line widths, with 81.5\% having FWZI $<10$~km~s$^{-1}$, whereas RSGs are preferentially distributed toward broader profiles, with 81.2\% having FWZI $>15$~km~s$^{-1}$. No RSG in the present sample is found at FWZI $<10$~km~s$^{-1}$. These results show that SiO maser line width is a useful empirical discriminator between AGB stars and RSGs. However, the separation is not complete, because a small number of AGB stars extend into the broad-line regime and overlap with the RSG distribution. {Checks of distance, maser transition, spectral resolution, and the use of the maximum FWZI among multiple available transitions do not remove this population-level difference. Sensitivity and maser brightness can affect the detectability of weak line wings, and multi-epoch AGB data show FWZI changes of a few km~s$^{-1}$; because an equivalent multi-epoch RSG sample is not available, a class-dependent variability bias cannot be completely excluded. FWZI should therefore be interpreted as a threshold-dependent empirical diagnostic rather than a bias-free intrinsic line width.}

\item We examined whether infrared colors provide a cleaner separation between SiO-maser-emitting AGB stars and RSGs. Both the IRAS two-color diagram and the [K$_\mathrm{s}$]$-$[W3] versus [K$_\mathrm{s}$]$-$[W4] diagram show broad statistical differences between the AGB and RSG populations. However, the separation is incomplete. In particular, known RSGs with SiO maser detections overlap substantially with the AGB SiO maser sample, and some broad-line AGB sources extend into color regions occupied by RSGs. Thus, infrared colors are useful qualitative diagnostics, but they do not provide a definitive classification criterion for SiO maser sources. In the present sample, infrared colors are not clearly superior to SiO maser line width as a discriminator between AGB stars and RSGs.

{\item We combined SiO maser FWZI with luminosities derived from infrared SED integration and adopted distances that are primarily RUWE-selected Gaia/Bailer-Jones estimates, with independent literature distances used for five RSGs. In the FWZI--luminosity plane, AGB stars are concentrated mainly at lower luminosities and narrower line widths, whereas RSGs tend to occupy broader-line and higher-luminosity regions. A weak positive trend is seen in the sense that sources with broader SiO maser profiles tend to have higher luminosities on average. A VLBI-based robustness check for the small subset with independent parallaxes does not reveal a large systematic displacement of the diagram. Nevertheless, the two populations still overlap substantially, and the combination of line width and luminosity does not yield a clean physical separation between AGB stars and RSGs. We therefore defined an empirical RSG-adjacent region, $15 \leq {\rm FWZI} \leq 26.6$~km~s$^{-1}$ and $3.9 \leq \log(L/L_\odot) \leq 4.9$, to characterize the high-luminosity and broad-line-width tail of the AGB-classified sample where the sources approach the RSG distribution most closely. Ten AGB-classified sources fall within this region and were selected for further analysis.}

\item For these ten selected sources, we examined additional diagnostics, including infrared SEDs and DUSTY fitting, OH and H$_2$O maser information, and period and variability information. Eight sources are broadly reproduced by simple one-dimensional single-shell DUSTY models. These satisfactory fits provide useful supporting constraints, but they do not {establish} a unique spherical circumstellar geometry. Two sources, J17423501$-$3005419 and J18420731$-$0423334, are not satisfactorily reproduced by the same simple model. Their poor SED fits indicate that the SED-based constraints are less reliable for these objects, or that their circumstellar environments may be more complex than assumed in the simple model. Possible causes include multiple dust components, uncertain extinction correction, photometric limitations, non-spherical circumstellar structure, or counterpart-identification problems. Period information from Gaia, AAVSO, and NEOWISE was also examined. NEOWISE W2 data are usable for only two of the ten sources, J17572314$-$1802456 and J18420731$-$0423334, and neither shows a secure period detection in the present analysis.

\item By combining the line width, luminosity, SED fitting, period information, and maser diagnostics, we tentatively divided the ten selected sources into three groups. Six sources, J17415458$-$2047055, J17572314$-$1802456, J19145678+2204305, J20561005+8303252, J21412708+7623113, and J22504945+6415047, are most naturally interpreted at present as luminous AGB-like objects. These sources enter the RSG-adjacent region because they are moderately broad-lined or relatively luminous, but their other diagnostics can still be understood within the range of ordinary or luminous AGB behavior. Two sources, J07463415$-$3218165 and J18085198$-$2616210, stand out as the objects for which follow-up in the high-initial-mass AGB, RSG-like, or SAGB-related context is most strongly motivated. They combine broad SiO maser FWZI, high luminosities, long periods, and outer-envelope maser activity. However, the present data are insufficient to classify them as RSGs or SAGB stars, or even as high-confidence SAGB candidates. The remaining two sources, J17423501$-$3005419 and J18420731$-$0423334, have poor DUSTY fits and must be interpreted more cautiously. Among them, J17423501$-$3005419 is especially noteworthy because it has high luminosity, broad SiO maser FWZI, a long period, and both OH and H$_2$O maser detections, whereas J18420731$-$0423334 lacks reliable period and OH/H$_2$O maser information and remains more uncertain.

\end{enumerate}

{Overall, our results show that SiO maser line width differs systematically between AGB stars and RSGs and can serve as a useful empirical diagnostic of evolved-star populations. However, line width alone, and even line width combined with luminosity, is not sufficient to determine the physical nature, detailed AGB evolutionary phase, or initial-mass regime of all sources. The RSG-adjacent region in the FWZI--luminosity plane should therefore be regarded not as a clean classification box for a specific type of object, but as a physically mixed parameter space that can isolate a small and astrophysically rich subset of the SiO maser population. This subset includes luminous AGB-like sources that may simply be at an advanced AGB phase, objects for which a high-initial-mass AGB/SAGB-related interpretation deserves follow-up, and sources with complex circumstellar environments. The FWZI--luminosity plane may therefore provide a useful screening tool for distinguishing these possibilities statistically and selecting targets for follow-up, rather than a definitive classifier. Further near-infrared spectroscopy, multi-epoch SiO maser monitoring, sensitive OH and H$_2$O maser observations, and molecular-line observations will be required to constrain both evolutionary phase and initial mass. Extending this approach to larger and well-curated SiO maser spectral databases, or to catalogs with uniformly measured FWZI values, will make it possible to test, in a more systematic way, how broad SiO maser kinematics, high luminosity, mass-loss properties, and evolved-star population membership are physically connected.}

\begin{acknowledgments}
This work is supported by the National SKA Program of China (No. 2025SKA0120100). The authors thank Professor Biwei Jiang of Beijing Normal University for facilitating the collaboration that led to this work and for her support throughout the project. The authors also thank Dr. Shuji Deguchi for kindly providing the SiO maser spectral data used in this study and for permitting the inclusion of the unpublished survey data. J.N. acknowledges support from the One Hundred Top Talent Program of Sun Yat-sen University (Grant No. 71000-18841229). Y.Z. acknowledges support from the National Natural Science Foundation of China (Grant Nos. 12473027 and 12333005) and the Guangdong Basic and Applied Basic Research Foundation (Grant No. 2024A1515010798). Z.Z. acknowledges support from the National Natural Science Foundation of China (Grant No. 125B2060). OpenAI's ChatGPT was used to assist with language editing, LaTeX/source consistency checks, reference and cross-reference checking, and final proofreading. All AI-assisted output was reviewed and verified by the authors, who take full responsibility for the scientific content and final wording of the manuscript.
\end{acknowledgments}

\appendix

\section{Unpublished SiO Maser Survey Observations}
\label{app:unpublished_sio_survey}

This appendix provides details on the unpublished SiO maser spectral data described in Section~\ref{subsec:data-used}. The target sources of the SiO maser survey presented here are primarily long-period variable stars selected based on the AAVSO International Variable Star Index \citep{2006SASS...25...47W} and the COBE DIRBE near-infrared light curve catalog \citep{2010ApJS..190..203P}. First, among the long-period variables listed in these two catalogs, we identified those that had not been previously observed for SiO maser lines.

Next, the sample was refined based on the following infrared criteria: $K_{s}<9$, $H-K_{s} \geq 0.4$, $F_{12} > 6\,\mathrm{Jy}$, and $-0.5 < \log(F_{25}/F_{12}) < 0.2$, where $F_{12}$ and $F_{25}$ denote the flux densities at 12\,$\mu$m and 25\,$\mu$m from IRAS, respectively. These criteria were designed to select oxygen-rich AGB candidates. The positional information for the SiO maser observations was obtained from infrared archival data.

The initial scientific objective of this SiO maser line survey was to investigate the kinematic properties of long-period variable stars from a Galactic dynamical perspective. Accordingly, proper motion criteria were also applied in the sample selection. Proper motion data were extracted from the PPMXL catalog \citep{2010AJ....139.2440R}. We selected sources whose PPMXL positions matched the 2MASS positions within $2''$ and satisfied the catalog-quality criterion $\mathrm{FL}=0$, where $\mathrm{FL}$ is the PPMXL quality flag and $\mathrm{FL}=0$ indicates no flag for excessively large coordinate $\chi^2$ or other catalog-quality issues. For cases with more than one possible PPMXL counterpart within the matching radius, we adopted the entry with the largest value of the PPMXL column $\mathrm{No}$, where $\mathrm{No}$ denotes the number of observations used in the PPMXL astrometric solution. Ultimately, 210 sources with reliable proper motion measurements were selected.

SiO maser line observations for these 210 targets were carried out from February to May 2011 using the Nobeyama 45-m radio telescope. The observed lines were the SiO $J=1\mbox{--}0$ transitions of $v=1$ and $v=2$, corresponding to 43.122\,GHz and 42.821\,GHz, respectively. A cooled HEMT receiver was used for the observations. The system temperature varied between 180 and 250\,K depending on atmospheric conditions. The half-power beam width (HPBW) at these frequencies was approximately $40''$. Observations were conducted in position-switching mode, and the flux density conversion factor was $2.9\,\mathrm{Jy\,K^{-1}}$. The velocity resolution was approximately $0.3\,\mathrm{km\,s^{-1}}$ for high-resolution scans and about $1.8\,\mathrm{km\,s^{-1}}$ for wideband scans. The observing procedures and data reduction followed the standard methods used at the Nobeyama 45-m telescope \citep{2000ApJS..128..571D}.

As a result of this survey, SiO masers were detected in 109 of the observed sources. Table~\ref{unpub_sio_sum} summarizes the maser line parameters of the 109 detected sources, including peak antenna temperatures and velocity-integrated line intensities. Table~\ref{unpub_sio_infrared} presents the basic properties of the 109 detected SiO maser sources, including variable-star information, near- and mid-infrared photometric quantities, IRAS counterparts, and representative SiO maser velocities.

\section{Notes on Individual Sources Selected for Further Analysis}
\label{app:individual_notes}

This appendix provides source-by-source notes for the ten sources selected for further analysis in Section~\ref{subsec:Integrated_interpretation}. For each source, we summarize the available FWZI, luminosity, period information, DUSTY-fit quality, SED and circumstellar-envelope characteristics, OH/H$_2$O maser information, and the main caveats that affect the interpretation. These notes serve as supporting material for the integrated interpretation presented in Section~\ref{subsec:Integrated_interpretation}. The FWZI, luminosity, and DUSTY-related quantities used here are summarized in Tables~\ref{tab:candidates} and~\ref{tab:dustyfits_full}. The period information is taken from external catalogs and literature sources, including Gaia DR3 LPV \citep{2023A&A...674A..15L} and the AAVSO International Variable Star Index \citep{2006SASS...25...47W}; the NEOWISE W2 light curves were examined using data from NEOWISE \citep{2014ApJ...792...30M}. The OH/H$_2$O maser information is compiled from the literature when available.

{For clarity, the H$_2$O detections discussed below refer to the 22.235~GHz line, and for every source with quoted OH information we specify whether the relevant measurement refers to the 1612~MHz satellite line or to the 1665/1667~MHz main lines.}

\paragraph{{J07463415$-$3218165} {--- Group 2 (high-initial-mass AGB/RSG-like/SAGB-related follow-up).}}
{This source has FWZI $\sim$26.6~km~s$^{-1}$ and $\log(L/L_{\odot}) \approx 4.88$. Period information is available from Gaia and AAVSO, with values of 844.5 and 725 days, respectively. The DUSTY fit is satisfactory, and the SED shows a {single-peaked} profile. {OH main-line maser emission at 1665/1667~MHz and 22.235~GHz H$_2$O maser emission have been reported. The OH profile quoted here refers to the main-line observations and shows emission features near 15 and 40~km~s$^{-1}$ \citep{1995A&AS..111..237L}. The 22.235~GHz H$_2$O maser spectrum displays peaks spanning a velocity range from 20.0 to 35.0~km~s$^{-1}$ \citep{2010ApJS..188..209K}.} Taken together, the broad SiO profile, high luminosity, and relatively long period make it more natural to group this source with the more RSG-like sources; however, the present observables do not uniquely distinguish between an extreme or late-stage luminous AGB interpretation and a higher-initial-mass AGB/SAGB-related interpretation.}

\paragraph{{J17415458$-$2047055} {--- Group 1 (luminous AGB-like).}}
This source has FWZI $\sim$16.3~km~s$^{-1}$ and $\log(L/L_{\odot}) \approx 3.91$. Period information is available from Gaia and AAVSO, with periods of 452.9 and 446 days, respectively. The DUSTY fit is satisfactory, and the SED shows a {single-peaked} profile. {OH 1612~MHz satellite-line maser emission has been detected and shows double peaks at 13.0 and 43.0~km~s$^{-1}$ \citep{1993A&A...277..453D}.} No H$_2$O maser information is currently available. Overall, this source is broadly consistent with an AGB-like source.

\paragraph{{J17423501$-$3005419} {--- Group 3 (poorly constrained SED/DUSTY fit).}}
This source has FWZI $\sim$23.0~km~s$^{-1}$ and $\log(L/L_{\odot}) \approx 4.83$. A Gaia period of 760.1 days is available, while no AAVSO period is available and no usable NEOWISE W2 light curve is available. The adopted line-of-sight extinction is extremely large, and the DUSTY fit is not reliable. {OH main-line maser emission at 1665/1667~MHz and 22.235~GHz H$_2$O maser emission have been reported. The OH profile quoted here refers to the main-line observations and shows emission features near $-30.0$ and 30.0~km~s$^{-1}$ \citep{1995A&AS..111..237L}, while the 22.235~GHz H$_2$O maser spectrum displays peaks spanning a velocity range from $-20.0$ to 0.0~km~s$^{-1}$ \citep{2010ApJS..188..209K}.} The poor SED fit may reflect the large adopted extinction, uncertainty in the extinction correction, photometric limitations, or counterpart-identification issues; a complex circumstellar environment is another possible explanation. This source is therefore difficult to classify from the present SED constraints alone, but it remains an astrophysically interesting broad-line, high-luminosity, maser-active object that deserves follow-up observations.

\paragraph{{J17572314$-$1802456} {--- Group 1 (luminous AGB-like).}}
This source has FWZI $\sim$18.5~km~s$^{-1}$ and $\log(L/L_{\odot}) \approx 4.70$. No reliable Gaia or AAVSO period is available. Usable NEOWISE W2 data are available, but no significant periodic signal was detected. The DUSTY fit is satisfactory, and the SED shows a {single-peaked} profile. No OH/H$_2$O maser information is currently available. The source is broadly consistent with an AGB-like source, although the broad line width leaves some room for further investigation.

\paragraph{{J18085198$-$2616210} {--- Group 2 (high-initial-mass AGB/RSG-like/SAGB-related follow-up).}}
This source has FWZI $\sim$22.9~km~s$^{-1}$ and $\log(L/L_{\odot}) \approx 4.85$. An AAVSO period of 720 days is available, while no Gaia period is available and no usable NEOWISE W2 light curve is available. The DUSTY fit is satisfactory, and the SED shows a {single-peaked} profile. {OH 1612~MHz satellite-line maser emission has been detected at approximately $-7.7$ and 29.7~km~s$^{-1}$ \citep{1991A&AS...90..327T}; the survey catalog reports the line parameters but does not present an individual spectrum for this source.} No H$_2$O maser information is currently available. Taken together, the available properties {make it more natural to group this source with the more RSG-like sources}, but a luminous AGB star cannot be ruled out on the basis of the current data.

\paragraph{{J18420731$-$0423334} {--- Group 3 (poorly constrained SED/DUSTY fit).}}
This source has FWZI $\sim$22.2~km~s$^{-1}$ and $\log(L/L_{\odot}) \approx 4.28$. No reliable Gaia or AAVSO period is available. Usable NEOWISE W2 data are available, but no significant periodic signal was detected. The DUSTY fit is not reliable. No OH/H$_2$O maser information is currently available. The source may be affected by a complex or highly obscured circumstellar environment, but alternative explanations cannot be excluded. This source therefore remains difficult to classify.

\paragraph{{J19145678+2204305} {--- Group 1 (luminous AGB-like).}}
This source has FWZI $\sim$16.1~km~s$^{-1}$ and $\log(L/L_{\odot}) \approx 4.14$. An AAVSO period of 371 days is available, while no Gaia period is available and no usable NEOWISE W2 light curve is available. The DUSTY fit is satisfactory, and the SED shows a {single-peaked} profile. No OH/H$_2$O maser information is currently available. Overall, this source is broadly consistent with an AGB-like source.

\paragraph{{J20561005+8303252} {--- Group 1 (luminous AGB-like).}}
This source has FWZI $\sim$22.2~km~s$^{-1}$ and $\log(L/L_{\odot}) \approx 4.16$. An AAVSO period of 535.2 days is available, while no Gaia period is available and no usable NEOWISE W2 light curve is available. The DUSTY fit is satisfactory, and the SED shows a {single-peaked} profile. No OH/H$_2$O maser information is currently available. Overall, this source is broadly consistent with an AGB-like source, although the broad line width leaves some room for further investigation.

\paragraph{{J21412708+7623113} {--- Group 1 (luminous AGB-like).}}
This source has FWZI $\sim$20.3~km~s$^{-1}$ and $\log(L/L_{\odot}) \approx 3.95$. An AAVSO period of 333 days is available, while no Gaia period is available and no usable NEOWISE W2 light curve is available. The DUSTY fit is satisfactory, and the SED shows a {single-peaked} profile. {H$_2$O maser emission at 22.235~GHz has been reported \citep{2013AJ....145...22K}, and the spectrum shows a {single-peaked} profile spanning from $-56.0$ to $-44.5$~km~s$^{-1}$.} No OH maser information is currently available. Overall, this source is broadly consistent with an AGB-like source.

\paragraph{{J22504945+6415047} {--- Group 1 (luminous AGB-like).}}
This source has FWZI $\sim$16.7~km~s$^{-1}$ and $\log(L/L_{\odot}) \approx 4.69$. An AAVSO period of 537 days is available, while no Gaia period is available and no usable NEOWISE W2 light curve is available. The DUSTY fit is satisfactory, and the SED shows a {single-peaked} profile. No OH/H$_2$O maser information is currently available. Overall, this source is broadly consistent with an AGB-like source, although the relatively high luminosity makes it worth retaining as a follow-up target.

\nocite{1997ApJ...478..206S}
\bibliography{ms}{}
\bibliographystyle{aasjournal}

\clearpage

\begin{deluxetable}{llc}
\tablecolumns{3}
\tablewidth{0pt}
\tablecaption{SiO maser spectral data summary\label{tab:sio_summary}}
\tablehead{
\colhead{Reference} & \colhead{Transition} & \colhead{Detections}
}
\startdata
\multirow{2}{*}{Shiki et al. (1997)} & $J=1$--$0$, $v=1$ & 4 \\
 & $J=1$--$0$, $v=2$ & 4 \\
\multirow{2}{*}{Izumiura et al. (1999)} & $J=1$--$0$, $v=1$ & 63 \\
 & $J=1$--$0$, $v=2$ & 66 \\
\multirow{2}{*}{Deguchi et al. (2000a)} & $J=1$--$0$, $v=1$ & 86 \\
 & $J=1$--$0$, $v=2$ & 85 \\
\multirow{2}{*}{Deguchi et al. (2000b)} & $J=1$--$0$, $v=1$ & 122 \\
 & $J=1$--$0$, $v=2$ & 122 \\
Deguchi et al. (2001) & $J=2$--$1$, $v=1$$^{\dagger}$ & 38 \\
\multirow{2}{*}{Ita et al. (2001)} & $J=1$--$0$, $v=1$ & 26 \\
 & $J=1$--$0$, $v=2$ & 24 \\
\multirow{2}{*}{Deguchi et al. (2004a)} & $J=1$--$0$, $v=1$ & 176 \\
 & $J=1$--$0$, $v=2$ & 168 \\
\multirow{2}{*}{Deguchi et al. (2004b)} & $J=1$--$0$, $v=1$ & 247 \\
 & $J=1$--$0$, $v=2$ & 251 \\
\multirow{2}{*}{Deguchi et al. (2005)} & $J=1$--$0$, $v=1$ & 43 \\
 & $J=1$--$0$, $v=2$ & 41 \\
\multirow{2}{*}{Nakashima \& Deguchi (2006)} & $J=1$--$0$, $v=1$ & 5 \\
 & $J=1$--$0$, $v=2$ & 2 \\
\multirow{2}{*}{Fujii et al. (2006)} & $J=1$--$0$, $v=1$ & 159 \\
 & $J=1$--$0$, $v=2$ & 152 \\
\multirow{4}{*}{Deguchi et al. (2007a)} & $J=1$--$0$, $v=0$ & 1 \\
 & $J=1$--$0$, $v=1$ & 2 \\
 & $J=1$--$0$, $v=2$ & 8 \\
 & $J=1$--$0$, $v=3$ & 1 \\
\multirow{2}{*}{Deguchi et al. (2007b)} & $J=1$--$0$, $v=1$ & 118 \\
 & $J=1$--$0$, $v=2$ & 113 \\
\multirow{2}{*}{Deguchi et al. (2010a)} & $J=1$--$0$, $v=1$ & 11 \\
 & $J=1$--$0$, $v=2$ & 7 \\
\multirow{2}{*}{Deguchi et al. (2010b)} & $J=1$--$0$, $v=1$ & 82 \\
 & $J=1$--$0$, $v=2$ & 80 \\
\multirow{4}{*}{Deguchi et al. (2012)} & $J=1$--$0$, $v=0$ & 2 \\
 & $J=1$--$0$, $v=1$ & 225 \\
 & $J=1$--$0$, $v=2$ & 223 \\
 & $J=1$--$0$, $v=3$ & 4 \\
\enddata
\tablecomments{
The last column gives the number of detections for each transition in the published surveys; these numbers are not unique-source counts. The unpublished survey presented in Appendix~\ref{app:unpublished_sio_survey} is not included in this table. The entries for the $J=1$--$0$, $v=0$ and $v=3$ transitions are listed for completeness only and are not used in the subsequent analysis. The $J=2$--$1$, $v=1$ observations marked with $\dagger$ were obtained with the Mopra 22-m telescope; the other published observations listed here were obtained with the Nobeyama 45-m telescope.
}
\end{deluxetable}

\begin{deluxetable}{lrrccc}
\tablecaption{RSGs with detected SiO maser emission used in this study}\label{tab:RSG_SiO}
\tablewidth{0pt}
\tablehead{
\colhead{Source Name} & 
\colhead{R.A. (J2016)} & 
\colhead{Dec. (J2016)} & 
\colhead{Cluster} & 
\colhead{Spectral Type$^{\star}$} & 
\colhead{References$^{\dagger}$} \\
\colhead{} & 
\colhead{(deg)} & 
\colhead{(deg)} & 
\colhead{} & 
\colhead{} & 
\colhead{}
}
\startdata
VY CMa                  & 110.74301 & $-$25.76754 & NGC 2362             & M2.5          & KIM10          \\
VX Sgr                  & 272.01685 & $-$22.22406 & $-$                  & M8.5          & KIM10          \\
NML Cyg                 & 311.60641 & 40.11652    & $-$                  & M7$-$M8       & KIM10          \\
S Per                   & 35.71546  & 58.58651    & $-$                  & M4.5          & KIM10          \\
MY Cep                  & 343.63207 & 60.82749    & NGC 7419             & M7.5          & KIM10          \\
$\mu$ Cep               & 325.87694 & 58.78003    & $-$                  & M1            & KIM10          \\
AH Sco                  & 257.82091 & $-$32.32519 & $-$                  & M5            & KIM10          \\
KY Cyg                  & 306.49189 & 38.35212    & NGC 6913             & M3 Ia         & KIM10          \\
IO Per                  & 46.69695  & 55.73316    & $-$                  & M3            & KIM13          \\
PZ Cas                  & 356.01364 & 61.78950    & $-$                  & M3            & KIM13          \\
UY Sct                  & 276.90222 & $-$12.46635 & $-$                  & M3.5          & KIM13          \\
V719 Cyg                & 300.91037 & 30.46928    & $-$                  & M4            & KIM13          \\
V386 Cep                & 343.30137 & 61.28344    & $-$                  & M3            & KIM13          \\
W Per                   & 42.65789  & 56.98341    & $-$                  & M4.5          & VER12          \\
IRC$-$10414             & 275.82461 & $-$13.71310 & $-$                  & M7            & IMA99, CHO12   \\
CD$-$33 12241           & 264.04216 & $-$33.49462 & Trumpler 27          & M0            & VER12          \\
2MASS J16470540$-$4550367 & 251.77252 & $-$45.84352 & Westerlund 1       & M2$-$M5 Ia    & FOK12          \\
2MASS J16470309$-$4552189 & 251.76294 & $-$45.87195 & Westerlund 1       & M3            & FOK12          \\
2MASS J18544742+0025265 & 283.69762 & 0.42404     & $-$                  & M4 I          & DEG04B         \\
2MASS J20211407+3537165 & 305.30860 & 35.62125    & $-$                  & M2.5 I        & DEG05          \\
2MASS J18375629$-$0652322 & 279.48455 & $-$6.87562 & RSGC1                & M3            & NAK06          \\
2MASS J18375528$-$0652484 & 279.48034 & $-$6.88012 & RSGC1                & M4            & NAK06          \\
2MASS J18375973$-$0653494 & 279.49892 & $-$6.89708 & RSGC1                & M4            & VER12          \\
2MASS J18375090$-$0653382 & 279.46209 & $-$6.89395 & RSGC1                & M0            & NAK06          \\
2MASS J18375890$-$0652321 & 279.49543 & $-$6.87559 & RSGC1                & M3            & NAK06          \\
2MASS J18203644$-$1602589 & 275.15184 & $-$16.04970 & [MCM2005b] 4         & $-$           & DEG10A         \\
2MASS J18283530$-$1056364 & 277.14709 & $-$10.94345 & [MCM2005b] 8         & $-$           & DEG10A         \\
2MASS J18285207$-$1057576 & 277.21697 & $-$10.96601 & [MCM2005b] 8         & $-$           & DEG10A         \\
2MASS J18285844$-$1056069 & 277.24351 & $-$10.93526 & [MCM2005b] 8         & $-$           & DEG10A         \\
2MASS J18385699$-$0606459 & 279.73747 & $-$6.11276 & RSGC2                & $-$           & DEG10A         \\
2MASS J18390238$-$0605106 & 279.75993 & $-$6.08628 & RSGC2                & $-$           & DEG10A, VER12  \\
2MASS J18390558$-$0604265 & 279.77326 & $-$6.07403 & RSGC2                & $-$           & DEG10A, VER12  \\
2MASS J18391961$-$0600408 & 279.83172 & $-$6.01134 & RSGC2                & $-$           & DEG10A, VER12  \\
2MASS J18392891$-$0556435 & 279.87047 & $-$5.94542 & RSGC2                & $-$           & DEG10A         \\
2MASS J18535249+0039313 & 283.46872 & 0.65870     & [MCM2005b] 13        & $-$           & DEG10A         \\
2MASS J18540749+0039039 & 283.53122 & 0.65109     & [MCM2005b] 13        & $-$           & DEG10A         \\
\enddata
\tablecomments{A dash in the Cluster column indicates that no cluster association is given. \\
$^{\star}$ Spectral types of RSGs. A dash in this column indicates that no reliable spectral type is available. \\
$^{\dagger}$ References for SiO maser observations of RSGs: IMA99: \citet{1999PASJ...51..587I}; DEG04B: \citet{2004PASJ...56..765D}; DEG05: \citet{2005PASJ...57..933D}; NAK06: \citet{2006ApJ...647L.139N}; DEG10A: \citet{2010PASJ...62..391D}; KIM10: \citet{2010ApJS..188..209K}; CHO12: \citet{2012AJ....144..129C}; KIM13: \citet{2013AJ....145...22K}; VER12: \citet{Verheyen2012}; FOK12: \citet{2012ApJ...760...65F}. \\
This table lists all RSGs with detected SiO maser emission used in this study, including sources contained in our SiO maser spectral dataset and additional RSGs compiled from the literature.}
\end{deluxetable}

\begin{deluxetable*}{lcccccccc}
\tablecaption{AGB-classified Sources Selected from the Empirical RSG-adjacent Region and Available Period Information\label{tab:candidates}}
\tablewidth{0pt}
\tabletypesize{\small}
\tablehead{
\colhead{2MASS Name} &
\colhead{R.A.} &
\colhead{Dec.} &
\colhead{$\log(L/L_{\odot})$} &
\colhead{FWZI} &
\colhead{{NEOWISE W2}} &
\colhead{Gaia Period} &
\colhead{AAVSO Period} &
\colhead{{Group}} \\
\colhead{} &
\colhead{(J2016, deg)} &
\colhead{(J2016, deg)} &
\colhead{} &
\colhead{(km~s$^{-1}$)} &
\colhead{{Status}} &
\colhead{(days)} &
\colhead{(days)} &
\colhead{}
}
\startdata
J07463415$-$3218165 & 116.6423  & $-32.30459$ & 4.877 & 26.6 & no data      & 844.5   & 725.0 & {2} \\
J17415458$-$2047055 & 265.47743 & $-20.78487$ & 3.908 & 16.3 & no data      & 452.9   & 446.0 & {1} \\
J17423501$-$3005419 & 265.64593 & $-30.09496$ & 4.833 & 23.0 & no data      & 760.1   & no data & {3} \\
J17572314$-$1802456 & 269.34643 & $-18.04601$ & 4.703 & 18.5 & Non-periodic & no data & no data & {1} \\
J18085198$-$2616210 & 272.21658 & $-26.27254$ & 4.845 & 22.9 & no data      & no data & 720.0 & {2} \\
J18420731$-$0423334 & 280.52997 & $-4.39155$  & 4.276 & 22.2 & Non-periodic & no data & no data & {3} \\
J19145678+2204305   & 288.73658 & $22.07514$  & 4.145 & 16.1 & no data      & no data & 371.0 & {1} \\
J20561005+8303252   & 314.04187 & $83.05700$  & 4.159 & 22.2 & no data      & no data & 535.2 & {1} \\
J21412708+7623113   & 325.36284 & $76.38648$  & 3.952 & 20.3 & no data      & no data & 333.0 & {1} \\
J22504945+6415047   & 342.70604 & $64.25131$  & 4.692 & 16.7 & no data      & no data & 537.0 & {1} \\
\enddata
\tablecomments{
The sources listed here are the ten AGB-classified objects selected from the empirical RSG-adjacent region defined in Figure~\ref{fig:FWZI_vs_L}. {The group assignments follow Section~\ref{subsec:Integrated_interpretation}: Group~1 denotes luminous AGB-like sources, Group~2 denotes sources for which high-initial-mass AGB/RSG-like/SAGB-related follow-up is most strongly motivated, and Group~3 denotes sources with poorly constrained SED/DUSTY fits.} The NEOWISE W2 light curves were examined for all sources. The label ``Non-periodic'' indicates that usable NEOWISE W2 data are available but show irregular variability without a significant periodic signal, while ``no data'' indicates that no usable NEOWISE W2 photometry is available. Missing entries in the Gaia and AAVSO period columns are indicated by ``no data''.
}
\end{deluxetable*}

\begin{deluxetable*}{lcrccccccc}
\tablecaption{Basic Information and DUSTY-Related Parameters of the Selected Sources and the Comparison Object\label{tab:dustyfits_full}}
\tablewidth{0pt}
\tablehead{
\colhead{Source Name} & 
\colhead{R.A.} & 
\colhead{Dec.} & 
\colhead{$A_V$} & 
\colhead{$\tau_{2.2}$} & 
\colhead{$T_{\mathrm{eff}}$} & 
\colhead{$T_{\mathrm{d}}$} & 
\colhead{$\dot{M}_{\rm total}$} & 
\colhead{$\chi^2_{\rm fit}$} & \colhead{{Group}} \\
\colhead{} & 
\colhead{(J2016, deg)} & 
\colhead{(J2016, deg)} & 
\colhead{(mag)} & 
\colhead{} & 
\colhead{(K)} & 
\colhead{(K)} & 
\colhead{($M_\odot\,\mathrm{yr}^{-1}$)} & 
\colhead{} & 
\colhead{}
}
\startdata
J07463415$-$3218165 & 116.642300 & $-$32.304590 & 3.7 & 0.161 & 4000 & 1000 & $5.0\times10^{-6}$ & $4.53\times10^{-4}$ & {2} \\
J17415458$-$2047055 & 265.477430 & $-$20.784870 & 2.3 & 0.732 & 3600 & 900 & $1.5\times10^{-5}$ & $4.12\times10^{-4}$ & {1} \\
J17423501$-$3005419 & 265.645930 & $-$30.094960 & 112.8 & $\cdots$ & $\cdots$ & $\cdots$ & $\cdots$ & $\cdots$ & {3} \\
J17572314$-$1802456 & 269.346430 & $-$18.046010 & 3.2 & 0.480 & 3000 & 900 & $9.0\times10^{-6}$ & $4.31\times10^{-4}$ & {1} \\
J18085198$-$2616210 & 272.216579 & $-$26.272540 & 3.8 & 0.782 & 3000 & 900 & $1.3\times10^{-5}$ & $4.75\times10^{-4}$ & {2} \\
J18420731$-$0423334 & 280.529971 & $-$4.391546 & 58.2 & $\cdots$ & $\cdots$ & $\cdots$ & $\cdots$ & $\cdots$ & {3} \\
J19145678+2204305 & 288.736580 & 22.075140 & 2.3 & 0.379 & 2800 & 900 & $6.4\times10^{-6}$ & $4.62\times10^{-4}$ & {1} \\
J20561005+8303252 & 314.041870 & 83.057000 & 0.4 & 0.547 & 3000 & 900 & $1.0\times10^{-5}$ & $4.18\times10^{-4}$ & {1} \\
J21412708+7623113 & 325.362840 & 76.386480 & 2.0 & 0.530 & 3200 & 900 & $1.1\times10^{-5}$ & $4.32\times10^{-4}$ & {1} \\
J22504945+6415047 & 342.706040 & 64.251310 & 2.7 & 0.413 & 3800 & 1000 & $9.0\times10^{-6}$ & $4.78\times10^{-4}$ & {1} \\
MSX SMC 055 & 12.529946 & $-$73.523655 & 0.8 & 0.983 & 3600 & 550 & $3.0\times10^{-5}$ & $4.01\times10^{-4}$ & {--} \\
\enddata
\tablecomments{
The ten selected sources are listed together with the comparison object MSX SMC 055. {The Group column uses the same assignments as Table~\ref{tab:candidates} and Section~\ref{subsec:Integrated_interpretation}; MSX SMC 055 is included only as a comparison object and therefore has no group assignment.} {For MSX SMC 055, the DUSTY-derived parameters were calculated with the same modeling assumptions adopted for the Galactic sources to provide a uniform comparison. Because its SMC environment differs in metallicity, the listed mass-loss rate is not used for a direct quantitative comparison with the Galactic sources.} For J17423501$-$3005419 and J18420731$-$0423334, only $A_V$ is listed because it is the line-of-sight extinction value adopted for the SED correction; the DUSTY-derived parameters are omitted because no reliable single-shell fit was obtained. The quantities $\tau_{2.2}$, $T_{\mathrm{eff}}$, $T_{\mathrm{d}}$, $\dot{M}_{\rm total}$, and $\chi^2_{\rm fit}$ are DUSTY-related fitting results. Here, $\dot{M}_{\rm total}$ represents the total gas+dust mass-loss rate derived from the DUSTY radiatively driven wind model under the assumptions described in Section~\ref{subsec:mass-loss_estimation}. These mass-loss rates should therefore be regarded as model-dependent estimates rather than independently measured values.
}
\end{deluxetable*}

\clearpage

\startlongtable
\begin{deluxetable*}{cccccccccc}
\tablecaption{Observational parameters of sources with SiO maser detections in the unpublished survey}\label{unpub_sio_sum}
\tablehead{
\colhead{2MASS Name} & 
\multicolumn{4}{c}{SiO $J=1-0$, $v=1$ maser line} & 
\multicolumn{4}{c}{SiO $J=1-0$, $v=2$ maser line} & 
\colhead{Obs. Date} \\
\cmidrule(lr){2-5} \cmidrule(lr){6-9}
& $T_A$ & $V_{\rm LSR}$ & Int. intensity & $\mathrm{rms}$ & 
$T_A$ & $V_{\rm LSR}$ & Int. intensity & $\mathrm{rms}$ & \colhead{} \\
& (K) & (km~s$^{-1}$) & (K~km~s$^{-1}$) & (K) & 
(K) & (km~s$^{-1}$) & (K~km~s$^{-1}$) & (K) & (yymmdd) \\
\cmidrule(lr){2-5} \cmidrule(lr){6-9}
}
\startdata
00010244+3830145 & 0.262 & $-$32.5 & 0.965 & 0.050 & 0.158 & $-$34.4 & 0.527 & 0.047 & 110408 \\
00312643+8235424 & $-$ & $-$ & $-$ & 0.085 & 0.338 & $-$22.5 & 0.192 & 0.070 & 110226 \\
01152971+4043082 & 0.601 & $-$13.8 & 1.190 & 0.091 & 0.472 & $-$14.6 & 1.333 & 0.084 & 110408 \\
01161114+4144580 & 1.432 & $-$44.1 & 7.407 & 0.119 & 1.248 & $-$45.3 & 5.995 & 0.114 & 110408 \\
01173454+0855525 & 0.274 & 1.6 & 0.236 & 0.097 & 0.501 & 1.2 & 1.319 & 0.084 & 110416 \\
01263692+5509355 & $-$ & $-$ & $-$ & 0.086 & 0.313 & $-$25.1 & 0.518 & 0.067 & 110226 \\
02160009$-$2031108 & 0.273 & 4.9 & 0.483 & 0.057 & $-$ & $-$ & $-$ & 0.055 & 110416 \\
02442214$-$2912237 & 0.743 & 8.7 & 0.895 & 0.106 & 0.518 & 8.2 & 0.860 & 0.095 & 110416 \\
04020883$-$1344553 & 0.666 & 7.8 & 2.145 & 0.096 & 0.823 & 7.9 & 2.116 & 0.086 & 110416 \\
04175536$-$1830255 & 2.349 & 47.3 & 6.075 & 0.127 & 1.095 & 45.4 & 3.873 & 0.112 & 110416 \\
04355719+2830524 & 0.302 & $-$0.7 & 0.453 & 0.090 & 3.671 & $-$1.1 & 9.125 & 0.080 & 110428 \\
04414807$-$0757187 & 0.378 & 40.8 & 1.155 & 0.087 & 0.851 & 41.1 & 2.080 & 0.078 & 110322 \\
04422129+0652398 & 5.147 & 11.3 & 17.968 & 0.094 & 5.510 & 11.0 & 18.024 & 0.083 & 110416 \\
05270426+5007281 & 3.544 & $-$28.1 & 14.830 & 0.089 & 1.024 & $-$32.4 & 3.463 & 0.085 & 110309 \\
05315578+0736486 & 0.263 & $-$16.7 & 1.134 & 0.055 & 0.242 & $-$14.8 & 0.760 & 0.048 & 110318 \\
06030264+1622400 & 0.871 & $-$58.4 & 2.705 & 0.117 & 0.954 & $-$58.1 & 2.037 & 0.100 & 110322 \\
06032577+1343561 & 0.909 & $-$13.5 & 3.538 & 0.112 & 0.675 & $-$13.8 & 2.794 & 0.092 & 110322 \\
06061681$-$0705511 & 2.756 & $-$29.3 & 7.095 & 0.106 & 4.793 & $-$29.3 & 9.175 & 0.089 & 110314 \\
06331735+3135304 & 0.355 & $-$57.7 & 0.821 & 0.073 & 0.721 & $-$55.4 & 1.437 & 0.063 & 110416 \\
06351932+6405336 & 0.698 & $-$34.3 & 3.261 & 0.076 & 1.001 & $-$36.1 & 3.316 & 0.071 & 110309 \\
06470582+0800535 & 0.738 & 22.6 & 1.393 & 0.104 & $-$ & $-$ & $-$ & 0.105 & 110322 \\
07010238+1044185 & 0.237 & $-$18.1 & 0.674 & 0.059 & 0.136 & $-$14.6 & 0.334 & 0.051 & 110416 \\
07072185+2817559 & 0.903 & $-$54.8 & 1.923 & 0.123 & 1.506 & $-$54.5 & 6.560 & 0.145 & 110206 \\
07124861$-$0755407 & 1.100 & $-$12.2 & 3.221 & 0.063 & 0.582 & $-$13.1 & 1.268 & 0.130 & 110207 \\
07151973$-$2200468 & 1.238 & 51.9 & 3.782 & 0.117 & 2.526 & 52.7 & 5.264 & 0.115 & 110418 \\
07261350+3322204 & 0.581 & 32.3 & 2.812 & 0.075 & 0.673 & 36.7 & 2.977 & 0.075 & 110207 \\
07443608+1402571 & 0.637 & $-$9.8 & 3.216 & 0.102 & 0.427 & $-$10.3 & 1.318 & 0.087 & 110322 \\
07452744+3046414 & 2.304 & 15.5 & 8.045 & 0.092 & 1.634 & 15.8 & 6.929 & 0.087 & 110416 \\
07463829$-$2620359 & 4.492 & 83.2 & 14.099 & 0.135 & 2.080 & 80.8 & 10.178 & 0.127 & 110418 \\
07543720$-$2804582 & 0.902 & 68.6 & 1.895 & 0.134 & 0.764 & 67.7 & 1.452 & 0.127 & 110418 \\
07570936+2119099 & 0.418 & 108.4 & 0.509 & 0.107 & 0.775 & 108.9 & 1.172 & 0.089 & 110322 \\
08092795+1439519 & 0.545 & 8.8 & 1.891 & 0.109 & 0.446 & 9.3 & 0.437 & 0.095 & 110206 \\
08133824+1011498 & 0.434 & 13.2 & 2.213 & 0.120 & 0.548 & 9.6 & 1.799 & 0.104 & 110322 \\
08294115$-$0619073 & 3.242 & 27.3 & 4.674 & 0.109 & 1.380 & 27.4 & 1.969 & 0.094 & 110418 \\
08360700+1312317 & 0.385 & $-$28.5 & 0.817 & 0.080 & $-$ & $-$ & $-$ & 0.071 & 110322 \\
08402415$-$1449200 & 3.896 & 25.1 & 14.932 & 0.102 & 2.409 & 25.4 & 11.852 & 0.098 & 110418 \\
09433137$-$1834286 & 0.295 & 40.3 & 0.790 & 0.059 & $-$ & $-$ & $-$ & 0.050 & 110429 \\
10000193+2115444 & 0.992 & $-$25.1 & 1.869 & 0.079 & $-$ & $-$ & $-$ & 0.090 & 110207 \\
10511904$-$2837403 & 0.286 & 12.8 & 1.143 & 0.066 & 0.398 & 10.3 & 0.573 & 0.058 & 110429 \\
12471961+3822305 & 1.506 & $-$25.2 & 6.218 & 0.094 & 1.316 & $-$24.5 & 3.875 & 0.081 & 110428 \\
15171471+3621336 & 0.572 & 52.1 & 1.575 & 0.043 & 0.315 & 49.1 & 0.570 & 0.044 & 110127 \\
15322484+0338275 & 1.348 & 23.7 & 4.861 & 0.090 & 2.044 & 23.8 & 5.501 & 0.108 & 110203 \\
15332161$-$2711031 & 0.575 & 11.2 & 1.367 & 0.121 & 0.796 & 10.2 & 1.619 & 0.104 & 110422 \\
15464372+3819206 & 0.399 & 10.5 & 1.498 & 0.061 & 0.413 & 10.4 & 1.175 & 0.077 & 110127 \\
15553341$-$1251054 & 0.338 & $-$26.8 & 1.596 & 0.099 & 0.407 & $-$18.7 & 1.016 & 0.085 & 110422 \\
16064451+4842242 & 0.459 & $-$26.3 & 1.523 & 0.070 & 0.858 & $-$25.0 & 2.765 & 0.086 & 110127 \\
16212446$-$0742001 & 0.528 & $-$23.8 & 2.311 & 0.079 & $-$ & $-$ & $-$ & 0.071 & 110422 \\
16232443$-$1551568 & 0.533 & 12.9 & 1.716 & 0.081 & 0.272 & 11.9 & 0.822 & 0.069 & 110427 \\
16385186+1403583 & $-$ & $-$ & $-$ & 0.090 & 0.536 & $-$28.2 & 0.845 & 0.110 & 110228 \\
16575100$-$0245424 & 0.289 & $-$18.4 & 0.561 & 0.087 & 0.543 & $-$18.9 & 0.604 & 0.104 & 110228 \\
17092875+1840501 & 1.240 & 28.1 & 3.633 & 0.128 & 1.079 & 27.9 & 5.410 & 0.150 & 110228 \\
17143820$-$0047202 & 0.584 & 35.0 & 1.784 & 0.079 & 0.522 & 34.9 & 0.987 & 0.070 & 110427 \\
17162484+0443166 & 1.507 & $-$65.7 & 4.254 & 0.114 & 0.708 & $-$66.0 & 3.372 & 0.111 & 110422 \\
17175047$-$1159331 & 1.567 & 5.0 & 4.085 & 0.134 & 1.005 & 5.8 & 2.193 & 0.118 & 110513 \\
17302294+1754505 & 0.921 & $-$52.6 & 4.357 & 0.123 & 1.081 & $-$57.1 & 1.684 & 0.120 & 110422 \\
17382760+0439421 & 0.713 & 46.0 & 2.076 & 0.122 & 0.948 & 46.6 & 2.697 & 0.101 & 110422 \\
17404520$-$1007296 & 1.691 & $-$35.4 & 7.403 & 0.117 & 0.668 & $-$34.1 & 2.902 & 0.102 & 110422 \\
17441010$-$0322203 & 1.056 & $-$199.1 & 2.287 & 0.081 & 1.276 & $-$198.9 & 2.248 & 0.082 & 110427 \\
17490172+2232230 & 0.468 & 17.3 & 1.328 & 0.118 & 1.150 & 16.9 & 4.223 & 0.110 & 110422 \\
17525583$-$2932111 & 0.422 & $-$32.0 & 1.076 & 0.065 & 0.388 & $-$26.2 & 0.845 & 0.075 & 110127 \\
17570307$-$1920159 & 0.532 & 56.9 & 0.819 & 0.104 & $-$ & $-$ & $-$ & 0.101 & 110422 \\
17571974+2946286 & 2.726 & $-$35.3 & 13.707 & 0.157 & 2.086 & $-$35.1 & 9.239 & 0.137 & 110422 \\
18535283$-$2222039 & 0.973 & 72.7 & 3.122 & 0.080 & 0.948 & 74.4 & 1.409 & 0.090 & 110127 \\
19131555+4658558 & 1.667 & $-$35.5 & 5.308 & 0.097 & 0.260 & $-$34.9 & 0.671 & 0.083 & 110112 \\
19291209+2216134 & 1.593 & 50.6 & 4.846 & 0.143 & 2.310 & 49.1 & 7.421 & 0.143 & 110513 \\
19400329+4353559 & $-$ & $-$ & $-$ & 0.063 & 0.427 & 33.7 & 0.969 & 0.069 & 110304 \\
19405238+4847414 & 2.694 & $-$22.7 & 4.564 & 0.107 & 1.796 & $-$22.7 & 3.067 & 0.116 & 110228 \\
19405904+3643328 & 0.856 & 16.0 & 4.258 & 0.080 & 0.505 & 12.0 & 2.322 & 0.100 & 110228 \\
19433776+4846413 & 1.670 & $-$101.7 & 3.214 & 0.103 & 1.019 & $-$101.6 & 2.186 & 0.118 & 110304 \\
19443800$-$1345063 & $-$ & $-$ & $-$ & 0.078 & 0.642 & 34.4 & 2.471 & 0.074 & 110516 \\
19491235+3024433 & 1.546 & 34.0 & 7.211 & 0.127 & 1.728 & 34.6 & 8.270 & 0.119 & 110513 \\
19512980+0427516 & 0.275 & 35.8 & 0.791 & 0.073 & 0.377 & 36.6 & 1.391 & 0.062 & 110516 \\
19514274+1624590 & 0.604 & 66.0 & 1.569 & 0.128 & 0.550 & 66.2 & 2.231 & 0.126 & 110513 \\
19590658$-$0753025 & 7.517 & 16.1 & 18.171 & 0.090 & 8.374 & 15.9 & 18.052 & 0.088 & 110516 \\
19592317+0006104 & 3.725 & $-$9.0 & 12.918 & 0.086 & 2.355 & $-$8.9 & 7.610 & 0.081 & 110516 \\
20062308+3425519 & 0.652 & $-$12.3 & 0.844 & 0.090 & 1.187 & $-$12.0 & 1.891 & 0.080 & 110408 \\ 
20063247+4322061 & 0.269 & $-$4.5 & 0.451 & 0.070 & 0.558 & $-$4.6 & 1.481 & 0.077 & 110228 \\
20143907+0447047 & 0.571 & 25.1 & 1.540 & 0.077 & 0.693 & 24.7 & 1.562 & 0.071 & 110517 \\
20191499+3831008 & 1.141 & $-$12.5 & 4.257 & 0.114 & 1.395 & $-$14.0 & 5.373 & 0.110 & 110408 \\
20214262+1717177 & 0.557 & 30.6 & 1.909 & 0.077 & 0.308 & 31.6 & 1.222 & 0.073 & 110517 \\
20265669+1506019 & 2.651 & 85.1 & 8.432 & 0.093 & 1.064 & 85.1 & 5.924 & 0.088 & 110517 \\
20290551+3942458 & 6.730 & $-$17.9 & 26.264 & 0.082 & 6.436 & $-$17.9 & 20.818 & 0.096 & 110304 \\
20323347+5457003 & 0.822 & $-$0.8 & 2.055 & 0.091 & 1.400 & $-$0.9 & 3.282 & 0.093 & 110327 \\
20323409$-$2141256 & 1.362 & 7.1 & 4.025 & 0.092 & 0.416 & 5.0 & 1.392 & 0.079 & 110214 \\
20350941+2834208 & 1.024 & 4.0 & 2.674 & 0.112 & 3.147 & 3.8 & 5.017 & 0.104 & 110408 \\
20413898+1152379 & 0.268 & $-$3.1 & 0.275 & 0.056 & $-$ & $-$ & $-$ & 0.052 & 110517 \\
20421889$-$0815307 & 1.039 & $-$51.3 & 3.033 & 0.093 & 0.947 & $-$51.5 & 2.244 & 0.090 & 110516 \\
20435841+1914170 & 0.260 & $-$25.8 & 0.635 & 0.087 & 1.403 & $-$25.5 & 3.363 & 0.080 & 110517 \\
20490085+3150558 & 5.480 & $-$60.1 & 22.415 & 0.169 & 5.282 & $-$62.2 & 26.088 & 0.157 & 110513 \\
20515317+4721204 & 0.396 & $-$13.5 & 1.781 & 0.090 & 2.225 & $-$13.6 & 7.885 & 0.090 & 110327 \\
21022076$-$2705148 & 1.624 & $-$52.8 & 3.970 & 0.087 & 1.093 & $-$53.0 & 2.846 & 0.094 & 110303 \\
21094663+1824492 & 0.393 & 29.8 & 1.745 & 0.067 & 0.355 & 29.1 & 1.579 & 0.060 & 110517 \\
21131822+3203271 & 0.360 & 31.0 & 1.738 & 0.068 & 0.260 & 29.1 & 0.479 & 0.063 & 110327 \\
21210620+2421378 & 1.027 & $-$18.5 & 4.263 & 0.082 & 2.553 & $-$18.3 & 7.225 & 0.079 & 110327 \\
21293265+4721343 & 0.855 & $-$6.4 & 3.797 & 0.088 & 0.652 & $-$6.1 & 2.599 & 0.099 & 110303 \\
21412708+7623113 & 2.156 & $-$54.6 & 15.561 & 0.093 & 0.929 & $-$47.0 & 4.983 & 0.088 & 110319 \\
21423496$-$3212102 & 0.676 & $-$31.6 & 1.824 & 0.082 & 0.406 & $-$28.4 & 2.284 & 0.100 & 110303 \\
21484218+4414582 & 0.624 & $-$14.7 & 1.355 & 0.066 & $-$ & $-$ & $-$ & 0.065 & 110516 \\
21550806+4813478 & $-$ & $-$ & $-$ & 0.054 & 0.172 & 4.0 & 0.398 & 0.042 & 110302 \\
21583715+4146245 & 3.273 & 19.7 & 12.161 & 0.088 & 2.018 & 19.9 & 6.746 & 0.105 & 110304 \\
21593341+4832153 & 0.208 & $-$10.5 & 0.652 & 0.049 & 0.288 & $-$9.6 & 0.663 & 0.047 & 110327 \\
22034583$-$2803038 & 0.438 & $-$98.4 & 0.729 & 0.102 & 0.710 & $-$99.7 & 1.089 & 0.082 & 110302 \\
22065167+4827557 & 0.440 & $-$11.9 & 0.885 & 0.064 & 0.322 & $-$12.6 & 0.461 & 0.058 & 110516 \\
22072984+3744069 & 0.480 & 1.9 & 1.815 & 0.066 & 0.508 & 1.5 & 2.096 & 0.061 & 110516 \\
22290091+4018556 & 0.513 & $-$52.2 & 1.374 & 0.097 & 1.001 & $-$48.0 & 1.258 & 0.083 & 110408 \\
22333249+5638563 & 0.452 & $-$27.1 & 1.216 & 0.082 & 0.704 & $-$27.3 & 1.935 & 0.096 & 110304 \\
22532037+3630262 & 0.407 & $-$59.9 & 1.398 & 0.079 & 0.623 & $-$56.4 & 1.391 & 0.074 & 110408 \\
23195703+6227483 & 0.434 & $-$73.0 & 1.980 & 0.059 & 0.561 & $-$75.5 & 2.413 & 0.070 & 110304 \\
23462432+5429085 & 1.432 & 16.8 & 5.669 & 0.091 & 0.289 & 16.8 & 0.926 & 0.090 & 110408 \\
\enddata
\tablecomments{
$T_A$ is the peak antenna temperature of the detected SiO maser line. $V_{\rm LSR}$ is the velocity with respect to the local standard of rest, measured at the velocity corresponding to the peak $T_A$ of the SiO maser emission. Int. intensity denotes the velocity-integrated line intensity, defined as $\int T_A\,dv$ (K km s$^{-1}$). $\mathrm{rms}$ gives the root-mean-square noise level measured from line-free channels, in K. A dash ($-$) indicates that no reliable value is available, typically due to non-detection or insufficient data quality. For non-detected transitions, $T_A$, $V_{\rm LSR}$, and Int. intensity are marked with a dash, while the $\mathrm{rms}$ noise level is still listed when available.
}
\end{deluxetable*}

\startlongtable
\begin{deluxetable}{lcccllrcr}
\tablecaption{Basic properties of sources with SiO maser detections in the unpublished survey}\label{unpub_sio_infrared}
\tablehead{
\colhead{2MASS Name} & 
\colhead{Var. Type} & 
\colhead{Period} & 
\colhead{$K_{s}$} & 
\colhead{$H-K_{s}$} & 
\colhead{IRAS Name} & 
\colhead{$F_{12}$} & 
\colhead{$\log(F_{25}/F_{12})$} & 
\colhead{$V_{\rm LSR}$} \\
\colhead{} & 
\colhead{} & 
\colhead{(days)} & 
\colhead{(mag.)} & 
\colhead{} & 
\colhead{} & 
\colhead{(Jy)} & 
\colhead{} & 
\colhead{(km s$^{-1}$)}
}
\startdata
00010244$+$3830145 & SR & 318.0 & 2.083 & 0.547 & 23584$+$3813 & 19.8 & $-0.352$ & $-33.5$ \\
00312643$+$8235424 & SR & 166.0 & 1.563 & 0.459 & 00278$+$8219 & 25.0 & $-0.410$ & $-22.5$ \\
01152971$+$4043082 & M & 347.7 & 2.657 & 0.530 & 01126$+$4027 & 13.4 & $-0.349$ & $-14.2$ \\
01161114$+$4144580 & M & 316.0 & 2.738 & 0.610 & 01133$+$4129 & 16.6 & $-0.345$ & $-44.7$ \\
01173454$+$0855525 & M & 404.6 & 1.820 & 0.476 & 01149$+$0840 & 29.8 & $-0.360$ & $1.4$ \\
01263692$+$5509355 & SR & 196.7 & 1.735 & 0.614 & 01234$+$5454 & 21.7 & $-0.280$ & $-25.1$ \\
02160009$-$2031108 & M & 364.0 & 2.613 & 0.502 & 02136$-$2045 & 26.7 & $-0.196$ & $4.9$ \\
02442214$-$2912237 & SRA & 277.0 & 1.469 & 0.427 & 02422$-$2925 & 21.8 & $-0.373$ & $8.4$ \\
04020883$-$1344553 & M & 400.0 & 2.802 & 0.472 & 03598$-$1353 & 24.7 & $-0.314$ & $7.9$ \\
04175536$-$1830255 & M & 299.0 & 1.175 & 0.408 & 04157$-$1837 & 42.4 & $-0.387$ & $46.3$ \\
04355719$+$2830524 & SRA & 418.0 & 1.568 & 0.678 & 04328$+$2824 & 54.6 & $-0.389$ & $-0.9$ \\
04414807$-$0757187 & M & 277.0 & 2.338 & 0.481 & 04393$-$0802 & 19.5 & $-0.487$ & $40.9$ \\
04422129$+$0652398 & M & 412.0 & 1.612 & 0.453 & 04396$+$0647 & 49.3 & $-0.213$ & $11.1$ \\
05270426$+$5007281 & M & 311.0 & 1.940 & 0.669 & 05231$+$5004 & 29.6 & $-0.310$ & $-30.2$ \\
05315578$+$0736486 & M & 331.0 & 1.746 & 0.400 & 05292$+$0734 & 24.8 & $-0.448$ & $-15.7$ \\
06030264$+$1622400 & M & 251.8 & 2.729 & 0.521 & 06001$+$1622 & 12.4 & $-0.425$ & $-58.3$ \\
06032577$+$1343561 & M & 415.2 & 2.939 & 0.591 & 06005$+$1344 & 18.9 & $-0.371$ & $-13.6$ \\
06061681$-$0705511 & M & 332.0 & 3.124 & 0.770 & 06038$-$0705 & 18.3 & $-0.386$ & $-29.3$ \\
06331735$+$3135304 & M & 385.7 & 2.419 & 0.563 & 06300$+$3137 & 28.5 & $-0.406$ & $-56.6$ \\
06351932$+$6405336 & M & 366.3 & 1.783 & 0.593 & 06304$+$6407 & 21.8 & $-0.407$ & $-35.2$ \\
06470582$+$0800535 & M & 397.5 & 2.820 & 0.464 & 06443$+$0804 & 10.7 & $-0.451$ & $22.6$ \\
07010238$+$1044185 & M & 430.3 & 3.629 & 0.660 & 06582$+$1048 & 33.8 & $-0.336$ & $-16.4$ \\
07072185$+$2817559 & M & 356.3 & 2.907 & 0.838 & 07042$+$2822 & 20.6 & $-0.310$ & $-54.7$ \\
07124861$-$0755407 & M & 432.0 & 3.404 & 0.466 & 07103$-$0750 & 12.7 & $-0.312$ & $-12.6$ \\
07151973$-$2200468 & M & 423.8 & 3.033 & 0.501 & 07132$-$2155 & 11.7 & $-0.359$ & $52.3$ \\
07261350$+$3322204 & M & 384.0 & 1.924 & 0.629 & 07229$+$3328 & 34.5 & $-0.368$ & $34.5$ \\
07443608$+$1402571 & M & 410.0 & 3.815 & 0.569 & 07417$+$1410 & 10.0 & $-0.371$ & $-10.1$ \\
07452744$+$3046414 & M & 424.1 & 2.205 & 0.668 & 07422$+$3054 & 27.3 & $-0.368$ & $15.6$ \\
07463829$-$2620359 & M & 391.0 & 2.706 & 0.740 & 07445$-$2613 & 77.9 & $-0.322$ & $82.0$ \\
07543720$-$2804582 & M & 330.4 & 2.666 & 0.428 & 07525$-$2757 & 9.7 & $-0.419$ & $68.1$ \\
07570936$+$2119099 & M & 321.1 & 3.606 & 0.448 & 07542$+$2127 & 11.2 & $-0.230$ & $108.7$ \\
08092795$+$1439519 & M & 139.7 & 3.386 & 0.663 & 08066$+$1448 & 17.1 & $-0.354$ & $9.1$ \\
08133824$+$1011498 & M & 366.0 & 3.292 & 0.489 & 08108$+$1021 & 7.7 & $-0.413$ & $11.4$ \\
08294115$-$0619073 & SRB & 245.5 & 0.039 & 0.405 & 08272$-$0609 & 106.0 & $-0.409$ & $27.3$ \\
08360700$+$1312317 & M & 228.6 & 2.873 & 0.464 & 08333$+$1322 & 11.0 & $-0.407$ & $-28.5$ \\
08402415$-$1449200 & M & 448.0 & 2.813 & 0.635 & 08380$-$1438 & 44.0 & $-0.246$ & $25.2$ \\
09433137$-$1834286 & M & 317.0 & 2.789 & 0.492 & 09411$-$1820 & 9.7 & $-0.383$ & $40.3$ \\
10000193$+$2115444 & M & 273.4 & 3.474 & 0.544 & 09572$+$2130 & 10.8 & $-0.421$ & $-25.1$ \\
10511904$-$2837403 & M & 338.6 & 2.621 & 0.405 & 10489$-$2821 & 13.8 & $-0.380$ & $11.6$ \\
12471961$+$3822305 & M & 345.6 & 2.833 & 0.780 & 12449$+$3838 & 22.8 & $-0.176$ & $-24.9$ \\
15171471$+$3621336 & M & 273.9 & 2.527 & 0.533 & 15152$+$3632 & 12.3 & $-0.376$ & $50.6$ \\
15322484$+$0338275 & M & 365.8 & 1.615 & 0.579 & 15298$+$0348 & 44.2 & $-0.364$ & $23.7$ \\
15332161$-$2711031 & M & 402.7 & 2.116 & 0.590 & 15303$-$2700 & 34.3 & $-0.418$ & $10.7$ \\
15464372$+$3819206 & SRB & 300.0 & 1.624 & 0.440 & 15448$+$3828 & 27.0 & $-0.398$ & $10.4$ \\
15553341$-$1251054 & M & 291.8 & 2.799 & 0.568 & 15528$-$1242 & 13.6 & $-0.398$ & $-22.7$ \\
16064451$+$4842242 & LB & 128.0 & 2.035 & 0.522 & 16052$+$4850 & 18.8 & $-0.354$ & $-25.6$ \\
16212446$-$0742001 & M & 332.7 & 2.243 & 0.488 & 16187$-$0734 & 22.4 & $-0.399$ & $-23.8$ \\
16232443$-$1551568 & M & 290.0 & 3.432 & 0.597 & 16205$-$1545 & 10.6 & $-0.387$ & $12.4$ \\
16385186$+$1403583 & M & 269.1 & 1.671 & 0.508 & 16365$+$1409 & 24.6 & $-0.450$ & $-28.2$ \\
16575100$-$0245424 & M & 180.6 & 2.627 & 0.493 & 16552$-$0241 & 13.0 & $-0.464$ & $-18.6$ \\
17092875$+$1840501 & M & 347.7 & 2.417 & 0.571 & 17072$+$1844 & 16.9 & $-0.337$ & $28.0$ \\
17143820$-$0047202 & M & 356.0 & 2.251 & 0.599 & 17120$-$0043 & 22.5 & $-0.413$ & $35.0$ \\
17162484$+$0443166 & M & 332.0 & 3.260 & 0.865 & 17139$+$0446 & 54.2 & $-0.398$ & $-65.8$ \\
17175047$-$1159331 & M & 269.9 & 3.529 & 0.671 & 17150$-$1156 & 20.8 & $-0.368$ & $5.4$ \\
17302294$+$1754505 & M & 263.7 & 3.246 & 0.568 & 17281$+$1757 & 12.3 & $-0.342$ & $-54.8$ \\
17382760$+$0439421 & M & 430.0 & 4.158 & 0.472 & 17359$+$0441 & 14.4 & $-0.183$ & $46.3$ \\
17404520$-$1007296 & M & 487.0 & 3.459 & 0.517 & 17379$-$1006 & 19.5 & $-0.334$ & $-34.7$ \\
17441010$-$0322203 & M & 277.6 & 3.186 & 0.514 & 17415$-$0321 & 19.2 & $-0.215$ & $-199.0$ \\
17490172$+$2232230 & M & 333.8 & 2.574 & 0.410 & 17469$+$2233 & 13.4 & $-0.377$ & $17.1$ \\
17525583$-$2932111 & M & 532.0 & 3.875 & 0.881 & 17496$-$2931 & 13.8 & $-0.370$ & $-29.1$ \\
17570307$-$1920159 & M & 401.5 & 1.617 & 0.593 & 17540$-$1919 & 83.9 & $-0.442$ & $56.9$ \\
17571974$+$2946286 & M & 399.0 & 2.239 & 0.548 & 17554$+$2946 & 24.2 & $-0.219$ & $-35.2$ \\
18535283$-$2222039 & M & 370.0 & 4.013 & 0.909 & 18508$-$2225 & 8.1 & $-0.352$ & $73.6$ \\
19131555$+$4658558 & M & 346.3 & 2.000 & 0.703 & 19118$+$4653 & 36.6 & $-0.368$ & $-35.2$ \\
19291209$+$2216134 & M & 233.0 & 4.059 & 0.727 & 19270$+$2209 & 17.4 & $-0.172$ & $49.8$ \\
19400329$+$4353559 & M & 366.7 & 1.452 & 0.513 & 19384$+$4346 & 31.2 & $-0.374$ & $33.7$ \\
19405238$+$4847414 & M & 422.0 & 2.316 & 0.687 & 19394$+$4840 & 27.6 & $-0.136$ & $-22.7$ \\
19405904$+$3643328 & M & 442.0 & 2.461 & 0.538 & 19391$+$3636 & 18.6 & $-0.325$ & $14.0$ \\
19433776$+$4846413 & M & 190.3 & 3.131 & 0.683 & 19422$+$4839 & 6.4 & $-0.462$ & $-101.6$ \\
19443800$-$1345063 & M & 363.4 & 3.106 & 0.525 & 19418$-$1352 & 14.6 & $-0.408$ & $34.4$ \\
19491235$+$3024433 & M & 330.8 & 2.527 & 0.670 & 19472$+$3017 & 19.9 & $-0.405$ & $34.3$ \\
19512980$+$0427516 & M & 350.0 & 2.796 & 0.507 & 19490$+$0420 & 12.0 & $-0.428$ & $36.2$ \\
19514274$+$1624590 & M & 388.0 & 4.229 & 0.681 & 19494$+$1617 & 10.7 & $-0.407$ & $66.1$ \\
19590658$-$0753025 & M & 421.0 & 2.301 & 0.585 & 19418$-$0750 & 106.0 & $-0.409$ & $16.0$ \\
19592317$+$0006104 & M & 335.0 & 4.514 & 1.173 & 19568$-$0002 & 17.4 & $-0.039$ & $-8.9$ \\
20062308$+$3425519 & M & 382.0 & 4.688 & 0.727 & 20044$+$3417 & 11.1 & $-0.210$ & $-12.1$ \\
20063247$+$4322061 & M & 174.0 & 3.280 & 0.667 & 20048$+$4313 & 13.0 & $-0.318$ & $-4.6$ \\
20143907$+$0447047 & M & 316.4 & 2.226 & 0.534 & 20121$+$0437 & 17.7 & $-0.360$ & $24.9$ \\
20191499$+$3831008 & M & 565.0 & 5.277 & 1.169 & 20174$+$3821 & 15.3 & $-0.139$ & $-13.2$ \\
20214262$+$1717177 & M & 409.0 & 3.759 & 0.683 & 20194$+$1707 & 15.8 & $-0.241$ & $31.1$ \\
20265669$+$1506019 & M & 526.3 & 4.265 & 0.688 & 20246$+$1456 & 15.4 & $-0.239$ & $85.1$ \\
20290551$+$3942458 & M & 770.0 & 3.750 & 1.030 & 20273$+$3932 & 30.6 & $-0.155$ & $-17.9$ \\
20323347$+$5457003 & M & 337.3 & 2.541 & 0.632 & 20312$+$5446 & 17.6 & $-0.441$ & $-0.9$ \\
20323409$-$2141256 & M & 347.4 & 2.983 & 0.525 & 20296$-$2151 & 32.1 & $-0.188$ & $6.1$ \\
20350941$+$2834208 & M & 425.4 & 2.779 & 0.705 & 20330$+$2823 & 23.5 & $-0.334$ & $3.9$ \\
20413898$+$1152379 & M & 468.4 & 2.543 & 0.568 & 20392$+$1141 & 36.9 & $-0.223$ & $-3.1$ \\
20421889$-$0815307 & M & 323.4 & 3.963 & 0.746 & 20396$-$0826 & 11.7 & $-0.103$ & $-51.4$ \\
20435841$+$1914170 & M & 482.0 & 1.699 & 0.527 & 20416$+$1903 & 47.3 & $-0.359$ & $-25.6$ \\
20490085$+$3150558 & M & 370.6 & 2.128 & 0.753 & 20469$+$3139 & 36.7 & $-0.372$ & $-61.2$ \\
20515317$+$4721204 & SRA & 275.7 & 0.590 & 0.473 & 20502$+$4709 & 105.0 & $-0.227$ & $-13.6$ \\
21022076$-$2705148 & M & 277.5 & 4.109 & 0.484 & 20593$-$2717 & 8.7 & $-0.371$ & $-52.9$ \\
21094663$+$1824492 & M & 398.0 & 3.578 & 0.550 & 21074$+$1812 & 14.1 & $-0.203$ & $29.5$ \\
21131822$+$3203271 & M & 297.0 & 4.652 & 0.756 & 21112$+$3150 & 10.9 & $-0.424$ & $30.0$ \\
21210620$+$2421378 & M & 409.0 & 2.889 & 0.572 & 21188$+$2408 & 19.1 & $-0.370$ & $-18.4$ \\
21293265$+$4721343 & M & 511.0 & 2.549 & 0.780 & 21277$+$4708 & 27.6 & $-0.347$ & $-6.2$ \\
21412708$+$7623113 & M & 333.0 & 1.397 & 0.777 & 21414$+$7609 & 47.6 & $-0.301$ & $-50.8$ \\
21423496$-$3212102 & M & 332.0 & 4.368 & 0.658 & 21396$-$3225 & 6.1 & $-0.269$ & $-30.0$ \\
21484218$+$4414582 & M & 304.5 & 3.017 & 0.763 & 21467$+$4400 & 16.3 & $-0.416$ & $-14.7$ \\
21550806$+$4813478 & M & 370.4 & 3.301 & 0.504 & 21532$+$4759 & 10.9 & $-0.428$ & $4.0$ \\
21583715$+$4146245 & M & 375.0 & 3.632 & 0.541 & 21565$+$4132 & 54.8 & $-0.210$ & $19.8$ \\
21593341$+$4832153 & M & 442.5 & 2.936 & 0.719 & 21576$+$4817 & 13.3 & $-0.427$ & $-10.1$ \\
22034583$-$2803038 & M & 271.0 & 3.965 & 0.504 & 22009$-$2817 & 8.8 & $-0.323$ & $-99.1$ \\
22065167$+$4827557 & M & 524.0 & 2.763 & 0.820 & 22049$+$4813 & 18.7 & $-0.330$ & $-12.2$ \\
22072984$+$3744069 & M & 328.5 & 2.541 & 0.510 & 22053$+$3729 & 15.0 & $-0.482$ & $1.7$ \\
22290091$+$4018556 & M & 241.5 & 2.257 & 0.609 & 22268$+$4003 & 21.3 & $-0.433$ & $-50.1$ \\
22333249$+$5638563 & M & 480.0 & 3.893 & 0.628 & 22316$+$5623 & 20.1 & $-0.306$ & $-27.2$ \\
22532037$+$3630262 & M & 366.3 & 3.556 & 0.532 & 22510$+$3614 & 9.4 & $-0.228$ & $-58.2$ \\
23195703$+$6227483 & M & 775.0 & 4.125 & 0.853 & 23177$+$6211 & 23.6 & $-0.194$ & $-74.2$ \\
23462432$+$5429085 & SRA & 415.0 & 1.332 & 0.621 & 23439$+$5412 & 39.4 & $-0.365$ & $16.8$ \\
\enddata
\tablecomments{
$V_{\rm LSR}$ is the representative SiO maser velocity, defined as the average peak velocity of the $v=1$ and $v=2$ transitions when both are detected, or as the peak velocity of the detected transition otherwise. $K_{s}$ and $H-K_{s}$ are taken from 2MASS, while $F_{12}$ and $F_{25}$ are taken from IRAS. Variability types and periods are taken from the latest AAVSO International Variable Star Index. Variability classes follow the standard nomenclature: M, Mira variables; SRA, semiregular variables with relatively stable periodicity; SRB, semiregular variables with poorly defined or multiple periodicities; SR, semiregular variables without further subclassification; and LB, slow irregular late-type variables.
}
\end{deluxetable}

\clearpage

\end{CJK*}
\end{document}